\documentclass[aps,rmp,twocolumn,amssymb,amsmath]{revtex4}
\usepackage{graphicx}
\usepackage{amsmath}
\usepackage{amssymb}
\usepackage{dcolumn}
\usepackage{color}

\def \be{\begin{equation}}
\def \ee{\end{equation}}
\def \ba{\begin{array}}
\def \ea{\end{array}}
\def \beq{\begin{eqnarray}}
\def \eeq{\end{eqnarray}}

\def \bed{\begin{displaymath}}
\def \eed{\end{displaymath}}

\begin{document}

\title{Nonequilibrium dynamics of closed interacting quantum systems.}

\author{Anatoli Polkovnikov$^1$, Krishnendu Sengupta$^{2}$, Alessandro Silva$^3$, Mukund Vengalattore$^4$ }
\affiliation {$^1$Department of Physics, Boston University, Boston, MA 02215, USA\\
$^2$ Theoretical Physics Department, Indian Association for the Cultivation of Science, Jadavpur, Kolkata 700 032, India\\
$^3$ The Abdus Salam International Centre for Theoretical Physics, Strada Costiera 11, 34151 Trieste, Italy\\
$^4$ Department of Physics, Cornell University, Ithaca, NY,
14853,USA}

\begin{abstract}
This colloquium gives an overview of recent theoretical and
experimental progress in the area of nonequilibrium dynamics of
isolated quantum systems. We particularly focus on quantum quenches:
the temporal evolution following a sudden or slow change of  the
coupling constants of the system Hamiltonian. We discuss several
aspects of the slow dynamics in driven systems and emphasize the
universality of such dynamics in gapless systems with specific focus
on dynamics near continuous quantum phase transitions. We also
review recent progress on understanding thermalization in closed
systems through the eigenstate thermalization hypothesis and discuss
relaxation in integrable systems. Finally we overview key
experiments probing quantum dynamics in cold atom systems and put
them in the context of our current theoretical understanding.
\end{abstract}

\maketitle

\tableofcontents

\section{Introduction}

In the past two decades the outlook of condensed matter physics has
been deeply and unexpectedly revolutionized by a few experimental
breakthroughs in atomic physics, quantum optics and nanoscience. In
synthesis, crucial advances in these fields have made it possible to
realize artificial systems (e.g. optical lattices, quantum dots,
nanowires) that are described to a very high degree of accuracy by
models (e.g. Hubbard, Kondo, and Luttinger models) whose physics has
been a subject of intense investigation in various contexts ranging
from high temperature superconductivity to low temperature transport
in metals. It is fair to say that this experimental progress has
changed the way theory and experiment look at each other. In the
past, effective models were largely devised to explain the low
energy physics of highly complex systems. The situation has now been
reversed so that one can experimentally realize and simulate the
physics of such models. On one hand, the design and realization of
interacting many-body systems could in principle be used to perform
practical tasks, such as quantum information
processing~\cite{farhi_01}. On the other hand, direct simulations of
simple models could help resolving important problems in condensed
matter physics. But most importantly, the availability of
experimental controllable systems whose properties can be accurately
described by simple models provides unprecedented opportunity to
explore several new frontiers of condensed matter physics including the
nonequilibrium dynamics in closed interacting quantum systems.

Equilibrium systems can often be understood using a combination of a
mean field theory, renormalization group, and universality. This
allows us to understand low temperature experimental data obtained
in complex systems, such as interacting electrons in solids, in
terms of simple effective models containing a few relevant
parameters. Away from equilibrium the situation is much less clear.
While some progress was made in the past for classical
systems ~\cite{Schmittmann_1995}, there are no rigorously justified
generalizations of any of these approaches to generic quantum
nonequilibrium systems. It is thus not obvious that the theoretical
study of the dynamics of simplified models would accurately describe
experiments of more complex systems. In addition there are fewer
available tools for analyzing dynamics of even simple interacting
models. In this respect cold atomic gases and
nanostructures make possible what would be arduous otherwise: a
fruitful comparison between nonequilibrium theories based on simple
models and carefully designed experiments with tunable
system parameters.

Finding systematic ways to understand the nonequilibrium physics of
interacting systems is not only of fundamental importance, but could
also be crucial for future technologies. A quantum computer, for
example, will definitely require the capability of performing
real time manipulations of interacting
quantum systems. Though large scale quantum computers are yet to be
on the horizon, it is evident that quantum coherent dynamics will be
one of the focus points of various experimental systems and of
future technologies.

Nonequilibrium dynamics is a potentially a vast field:
there are many ways to take a system out of equilibrium, such as
applying a driving field or pumping energy or particles in the
system through external reservoirs as in transport problems.
It is thus of utter importance to focus on simple, yet
fundamental protocols. In this Colloquium we will concentrate
on the simplest paradigm:  the study of the nonequilibrium
dynamics of closed interacting quantum systems following a change in
one of the system parameters (quantum quench). Such a change, which
could be either fast or slow, is particularly interesting when it
takes the system through a quantum phase transition  involving
macroscopic changes in the state of the many-body system at the
initial and the final point. Seminal work in this direction includes
groundbreaking experiments~\cite{greiner2002a, greiner2002b} showing both the feasibility
of observing a quantum phase transition in cold atoms and the
possibility of observing quantum coherent dynamics. Following this
work, a number of different experiments explored the dynamics of
cold atom systems driven across BCS-BEC crossover~~\cite{regal_thesis}, polar and ferromagnetic
phases of spinor condensates ~\cite{sadler_06}, insulating and
superfluid phases of ultracold bosons ~\cite{tuchman_06} and many
others (see Ref.~\cite{bloch_review} for a review).

These experiments stimulated an active theoretical research in the relatively unexplored area of quantum dynamics in closed interacting systems. An interesting characteristic common to these systems is that despite of the absence of energy exchange with an environment and of the consequent global relaxation, it nevertheless frequently possible to
look at the long time dynamics and characterize it in terms of an asymptotic state attained by
physical (measurable) observables~\cite{rigol_08, reimann_08, linden_2009, gogolin_11, cramer_08,flesch_08}. In connection to this, it is possible to categorize recent research on the subject of this Colloquium in two main questions\rm:
\begin{itemize}
\item What is universal in the dynamics of a system following a quantum quench~?
\item What are the characteristics of the asymptotic, steady state reached after a quench~? When is it thermal~?
 \end{itemize}

In this Colloquium we will discuss both of these questions
extensively. We shall outline our current level of understanding of
these issues and chart out the outstanding open questions in the
field. In Sec.~\ref{secII} we will focus on the first question and
describe, from various points of view, the universal aspects of
nearly adiabatic dynamics near quantum critical points as well as in
generic gapped and gapless systems. We will argue that the proximity
to the adiabatic limit allows us to make specific universal
predictions of scaling of various quantities such as the defect density and
heat with the quench rate.

In Sec.~\ref{secIII}, we will discuss recent progress in
understanding thermalization of a quantum system following a quench.
In classical systems active interest in this topic was stimulated by
the celebrated work of Fermi Pasta and Ulam on the dynamics of a
one-dimensional (1D) anharmonic chain~\cite{fpu, campbell_05,
porter_09} which demonstrated the absence of such thermalization. It was
realized much later that the nonlinearity of the interaction is not
sufficient for thermalization which occurs, in this system, only if
the initial amplitude of interaction exceeds a certain
threshold~\cite{izrailev_chirikov}. Below this threshold, the
solution splits into solitons and retains its quasi-periodic
nature~\cite{zabusky_kruskal} which is a consequence of the
Kolmogorov-Arnold-Moser (KAM) theorem~\cite{tabor}. In quantum
systems, the question of sufficient criteria for thermalization has
remained largely unaddressed so far. Some experimental progress in
this direction has been made by a recent experiment from Kinoshita
{\em et. al.}~\cite{kinoshita} on non-thermalizing dynamics of 1D
bosons with short range interactions. This experiment constitutes
the first clear demonstration of the fact that a nearly integrable
quantum interacting many-particle system does not thermalize for a
very long time. Currently, the question of extension of KAM theorem
to quantum systems is a subject of active theoretical debate~(see
e.g. Ref.~\cite{olshanii_09}).

Finally, we note that many important topics concerning the physics
of closed interacting systems did not find space in this Colloquium.
Most important among these are the tools that are being developed to
describe theoretically the physics of interacting systems out of
equilibrium. Among such methods we mention density-matrix renormalization group (DMRG) and time-evolving block decimation (TEBD) for analyzing equilibrium and nonequilibrium 1D systems ~\cite{white_92,
schollwoeck_05, schollwoeck_06, vidal_03, vidal_04} and higher
dimensional ones~\cite{verstrate_08}, the Keldysh
technique~\cite{kamenev_09} which is particularly
helpful for deriving quantum kinetic equations, and closely related
functional integral methods~\cite{plimak_01, gasenzer_09, rey_05}.
Cold atom experiments also prompted rapid developments in phase
space methods, where quantum dynamics is represented as an evolution
in the classical phase space~\cite{blakie_08, polkovnikov_09}. These
methods were originally developed and applied to various problems in
single-particle dynamics~\cite{hillery_84, zurek_03} and
independently in quantum optics in the context of coherent
states~\cite{walls-milburn, gardiner-zoller}. There are other
reviews available in literature (see the references above) which
specifically target these areas.

\section{Nearly adiabatic dynamics in quantum systems}
\label{secII}

\subsection{Universality in a nutshell}

Universality (or insensitivity to microscopic details) is one of the
crucial concepts of modern condensed matter physics. It naturally
emerged from one of the milestones of modern physics: the
renormalization group. In condensed matter physics universality is a
very powerful tool for the understanding of continuous (second
order) phase transitions, both classical~\cite{LL5,chaikin-lubensky}
and quantum~\cite{Sachdev_book, sondhi1997rmp, vojta2003progphys}.
As a consequence of the divergence of the correlation
length, a system undergoing such a continuous phase transition is
typically scale invariant in the vicinity of the critical point and
can be characterized by relatively simple massless field theories,
which permit a classification of perturbations driving the system
away from the critical point. Consequently, universality manifests
itself in the scaling behavior of various quantities such as the
order parameter, (free) energy, susceptibilities and correlations
functions near the critical point. In this review we will focus
mostly on quantum phase transitions occurring at zero temperature
upon the variation of a control parameter $\lambda$ through a
critical point $\lambda_c$. A standard example of universality is
the fact that the exponent $\nu$ characterizing the divergence of
the correlation length $\xi\sim 1/|\lambda-\lambda_c|^\nu$ near the
quantum critical point (QCP) is insensitive to the microscopic
details of the system and depends only on  the \it universality
class \rm of the transition, determined by the dimensionality,
overall symmetries and range of the interactions. For classical
(thermal) phase transitions similar universality manifests in the
divergence of the relaxation time $\tau_{rel}\sim
1/|\lambda-\lambda_c|^{z\nu}$, where $z$ is the dynamical critical
exponent. For quantum phase transitions the exponent $z$ can be
associated with a vanishing energy scale $\Delta\sim
|\lambda-\lambda_c|^{z\nu}$, which can be either a gap or a
crossover scale where the spectrum changes qualitatively. By the
uncertainty principle this energy scale corresponds to a divergent
time scale, which typically describes the crossover in the scaling
behavior of unequal time correlation functions. Phase transitions
can be also characterized by singular susceptibilities, which are in
turn connected through the fluctuation-dissipation theorem to the
correlation functions of conjugate variables (e.g., the magnetic
susceptibility is related to the correlation function of the
magnetization). At critical points these correlation functions have
often power law scaling behavior at long distances, e.g. $\langle
m(x) m(x')\rangle \approx 1/|x-x'|^{2\alpha}$. The exponent $\alpha$
sets the scaling dimension of the corresponding operator $m(x)$:
$\dim[m(x)]=\alpha$. Because similar correlation functions can enter
different susceptibilities not all the scaling exponents are
independent but must satisfy scaling relations
~\cite{chaikin-lubensky, vojta2003progphys}.

As mentioned in the introduction, the idea of universality makes it
possible to interpret experimental data obtained in real systems in
terms of effective models with a few parameters. Universality can be
ultimately understood using the renormalization group, which shows
that as a system is coarse grained to lower energies and longer
length scales, more and more parameters of its original, \it ab
initio \rm description become unimportant (irrelevant), while the
remaining few (relevant) parameters define an effective low energy
model. A standard example of universality in this context is the
scaling relation between energy and momentum of elementary
excitations, $\epsilon\propto k^z$, controlled by the dynamical
exponent $z$ which depends on the symmetries of the system. In
particular, $z=1$ in most phases characterized by a continuous
broken symmetry (crystals, superfluids, anti-ferromagnets), $z=2$ in
ferromagnets, where there is an additional conservation law of the
order parameter.

Universality is well established and understood in equilibrium. It
is, however, crucial for many experimentally relevant situations to
understand the extent to which this concept can be extended to out
of equilibrium physics. Can irrelevant interactions turn out to be important
away from equilibrium? Since there are many ways to take a system
out of equilibrium, for which specific protocols will universality
emerge and which details of the protocol are potentially important?
Below we will focus on recent studies addressing these
important issues in closed interacting quantum systems, and in
particular on the dynamics  of a system whose parameters are
dynamically tuned either through a quantum critical point, or in
general within a gapless/gapped phase.

\subsection{ Universal dynamics near quantum critical points}
\label{sec:Dynamics_QCP}

Let us start by considering the simplest nonequilibrium protocol~\cite{ap_adiabatic, zurek_05, dziarmaga_05}: the system is prepared in its ground state and is then driven through a QCP by
changing an external parameter $\lambda$ in time. As
long as the rate change of the gap in the spectrum
$\Delta$~\footnote{We use the word ``gap'' for brevity. However, the
system can be gapless on one or both sides of the transition (e.g.
superfluid-insulator transition). Then $\Delta$ would denote a
crossover energy scale vanishing at the QCP.} caused by changing
$\lambda$ is smaller than the square of the gap one can expect  the
system to approximately follow the ground state adiabatically (we
will revisit this statement in the next section). However, the
vanishing of the gap at $\lambda=\lambda_c$ implies that the system
will always violate adiabaticity close to the quantum critical
point, no matter how slowly the parameters are changed. It is then
natural to ask how many excitations will be generated while passing
though the critical point and how their density as
well as generated entropy and energy will depend on the rate of
change of $\lambda$.

A similar question known under the name of the Kibble-Zurek mechanism (KZ)~\cite{kibble, zurek1, zurek2, kibble2007} has been addressed in the past decades for classical phase transitions. In that case the excitation density of defects is described by a simple scaling argument~\cite{zurek1, zurek2}. Suppose that the tuning parameter, for example external temperature $T$, slowly decreases in time across the critical value $T_c$: $T=T_c-\upsilon
t$. The system will respond adiabatically (quasi-statically if the
system is not thermally insulated) up to some close vicinity of the
critical point, where adiabaticity will be violated as a result of
the divergence of the relaxation time ($\tau_{rel}\sim
1/|T-T_c|^{z\nu}$) and the dynamics will become diabatic (sudden).
The adiabatic response is once again resumed after the system moves
out of the vicinity of the critical point. Zurek
suggested a very simple criterion for separating such adiabatic and
diabatic (impulse) regimes: the time to reach the critical point
$t=|T-T_c|/\upsilon$ should be equal to the relaxation time. This
immediately introduces the time and length scales characterizing the
adiabatic to diabatic crossover: $t^\star\sim
1/|\upsilon|^{z\nu/(z\nu+1)}$, $\xi^\star\sim
1/|\upsilon|^{\nu/(z\nu+1)}$. The violation of adiabaticity implies
that  order can not form on distances larger than $\xi^\star$
leading to the formation of a domain structure with a characteristic
distance $\xi^\star$ between the domain boundaries.
\begin{figure} \includegraphics[width=\linewidth]{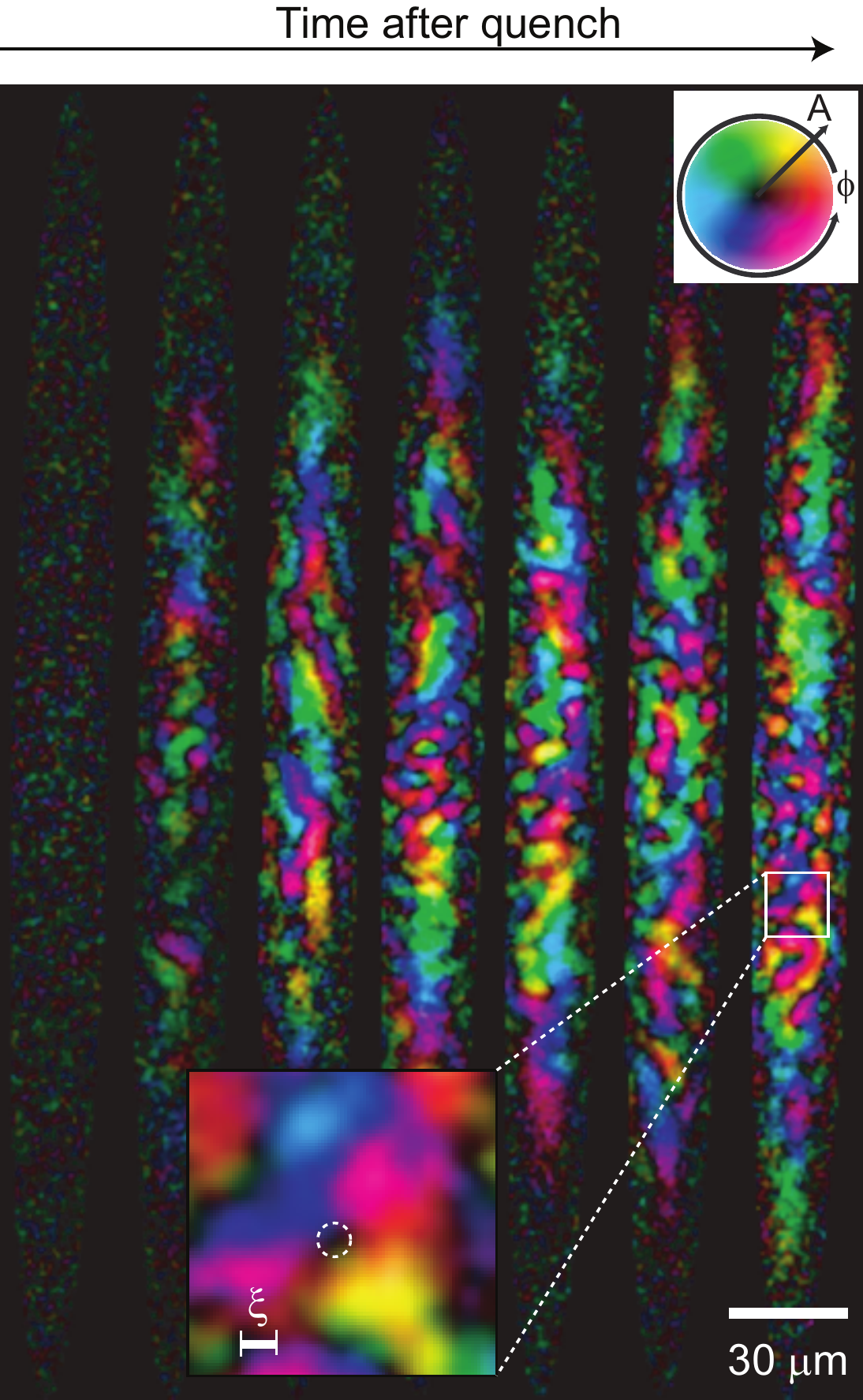}
\caption{{\bf Defect generation after a quench in a spinor condensate} -
These images show the transverse magnetization density of spinor condensates
for variable evolution times after a quench to a ferromagnetic state,
revealing a spatially inhomogeneous formation of ferromagnetic
domains. The orientation $\phi$ and amplitude $A$ are depicted by the color and brightness according to
the color wheel shown. Inset: An instance of a spin vortex spontaneously
created during the quench. For reference, the length scale corresponding to
the characteristic healing length $\xi$ is also shown. (Adapted from Ref.~\cite{sadler_06}. See Sec.~\ref{sec:exp} for more details.) }
\label{fig:KZ}
\end{figure}
In two and three dimensional systems, when the order parameter is
characterized by a continuous broken symmetry, the points where
several domains meet correspond to vortices or vortex rings. These are robust topological excitations with a very long
life time (see Fig.~\ref{fig:KZ}). Since $\xi^\star$ determines the
average distance between the defects their density is given by a
simple universal expression
\be n_{\rm ex}\sim
(\xi^\star)^d\sim|\upsilon|^{d\nu/(z\nu+1)}. \label{kz_zcaling}
\ee
The universality of the KZ prediction above is manifest in
the appearance of the universal critical exponents $z$ and $\nu$ in
the scaling law. This scaling was confirmed in experiments in liquid
crystals~\cite{ducci_1999}. Experiments in other systems
(superconductors~\cite{maniv_2003}, arrays of Josephson
junctions~\cite{monaco_2006}) observed the production of topological
defects with a power law scaling on the quench rate but gave a
different exponent. The KZ scaling was also confirmed
theoretically using stochastic dynamics (Ginzburg-Landau dynamics
with Langevin noise or Glauber dynamics) where temperature changes
in time~\cite{laguna_1997, yates_1998, krapivsky_2010, rivers_2001}, though there
are also works suggesting various modifications~\cite{biroli_10,
hindmarsh_2000}. One can also interpret Eq.(\ref{kz_zcaling}) as a
measure of non-adiabaticity near the critical point.
It is naturally expected that other measures like non-adiabatic
energy production and entropy generation will display similar
universality. These measures might be preferable over $n_{\rm ex}$
in situations where it is difficult to identify defects. Finding the
scaling of these quantities remains an open question.

The arguments above were recently generalized to the crossing of
quantum phase transitions~\cite{ap_adiabatic, zurek_05,
dziarmaga_05} (for recent reviews on this subject see
Refs.~\cite{dziarmaga_10, gritsev_10}. As discussed before in the
quantum case the parameter to be varied is not temperature $T$ but
rather the coupling $\lambda$ tuning the system through the quantum
critical point. In order to obtain the scaling for the number of
excitations produced in the quantum case let us first recall the
Landau-Zener analysis of the crossover between adiabatic and
nonadiabatic dynamics in a simple driven two-level system: \be
\mathcal H_{lz}=g(t)\sigma_z+\Delta\sigma_x. \label{lz_hamilt}, \ee
where $g(t)=\upsilon t$. If the system was initially prepared in the
ground state at $t\to-\infty$, the probability of transition to the
excited state at $t\to+\infty$ is~\cite{L, Z, M, S} \be p_{\rm
ex}=\exp[-\pi \gamma], \label{zl_prob} \ee where we introduced the
Landau-Zener parameter $\gamma=\Delta^2/\upsilon$. Notice that the
limit $\gamma\gg 1$ corresponds to the adiabatic limit with an
exponentially suppressed transition probability while $\gamma\ll 1$
corresponds to the diabatic limit where the transition happens with
probability close to unity. Hence when the rate of change of the
energy splitting between two levels becomes larger or comparable to
the energy splitting squared one observes a crossover from adiabatic
to diabatic dynamics. An alternative qualitative explanation of this
result has bee recently formulated ~\cite{damski_06}.

The Landau-Zener argument can be straightforwardly extended to the
crossing of a QCP. The characteristic energy scale which changes in
time is now the gap $\Delta$. As we discussed earlier this gap
universally depends on the tuning parameter $\lambda$:
$\Delta(\lambda)\sim |\lambda-\lambda_c|^{z\nu}\sim |\upsilon
t|^{z\nu}$, where we assumed that the dependence $\lambda(t)$ can be
linearized near the QCP: $\lambda(t)\approx \lambda_c+\upsilon t$.
Comparing the rate of change of the gap with its square, i.e.
solving the equation $d\Delta/dt\approx \Delta^2$, we find the
energy scale at which adiabaticity breaks down is $\Delta^\star\sim
|\upsilon|^{z\nu/(z\nu+1)}$. At this point the system is
characterized by the length scale $\xi^\star\sim
|\upsilon|^{-\nu/(z\nu+1)}$, which can be interpreted as the typical
length scale of fluctuations of the order parameter. Beyond this
point the adiabatic approximation breaks down and fluctuations at
longer scales cannot adiabatically follow the ground state. This
results in the creation of defects with typical distance $\xi$
between them and density $n_{\rm ex}\sim |\xi^\star|^d\sim
|\upsilon|^{d\nu/(z\nu+1)}$. This scaling is identical to the
classical one, Eq. ~(\ref{kz_zcaling}) with $\lambda \to T$, and was
proposed independently in Refs.~\cite{ap_adiabatic, zurek_05}. There
is a simple quasi-particle interpretation for this scaling: assuming
that the excitations in the system are characterized by isolated
quasi-particles then their density can be found from $n_{\rm
ex}\approx \int_0^{\Delta^\star} d\epsilon \rho(\epsilon)$, where
$\rho(\epsilon)$ is the single-particle density of states near the
QCP. In uniform d-dimensional systems $\rho(\epsilon)\sim
\epsilon^{d/z-1}$, which again reproduces Eq.~(\ref{kz_zcaling}).

The scaling in Eq.(\ref{kz_zcaling}) was verified in a series of
exact solutions of the dynamics across the QCP in integrable systems
whose dynamics can be mapped into a series of Landau-Zener
transitions of a few quasi-particle modes. In particular, it has
been verified for various spin models in one and two dimensions
which can be mapped to noninteracting
fermions~\cite{mukherjee_2007}, for models where low energy
excitations near phase transitions can be described by bosonic
Goldstone modes~\cite{ap_adiabatic, lamacraft_2007, dziarmaga_08},
the sine-Gordon model, where elementary excitations are solitons and
breathers with fractional statistics~\cite{degrandi_08,
degrandi_2010}, graphene~\cite{dora_10, dora_10a}. This scaling was
also extended to disordered systems, like a disordered Ising spin
chain, where it was found that $n_{\rm ex}\sim
1/\log^2(\upsilon)$~\cite{caneva_07, dziarmaga_06}, as expected from
Eq.~(\ref{kz_zcaling}) due to the divergence of the exponent $z$
near the critical point~\cite{fisher_95}.

The scaling in Eq.(\ref{kz_zcaling}) can be generalized to the case
of nonlinear dependence of the tuning parameter on time,
$\lambda(t)\sim \lambda_c\pm \upsilon |t|^r$, where considerations
similar to those leading to Eq.~(\ref{kz_zcaling})
give~\cite{sen_2008, optimal_passage, degrandi_10a}: \be n_{\rm
ex}\sim |\upsilon|^{d\nu/(z\nu r+1)}. \label{kz_zcaling1} \ee In all
cases $\upsilon$ in Eq.~(\ref{kz_zcaling1}) plays the role of the
adiabatic parameter: the limit $\upsilon\to 0$ corresponds to the
adiabatic limit (this interpretation is valid even for instantaneous
quenches $r=0$, where $\upsilon$ plays the role of the quench
amplitude). This suggests the dynamics  can be systematically
analyzed using  adiabatic perturbation theory ~\cite{ap_adiabatic,
ortiz_2008, degrandi_2009}, i.e. expanding the transition amplitudes
to the instantaneous eigenstates of the system in powers of
$\upsilon$. Using such analysis in Refs.~\cite{degrandi_10a, degrandi_2010} it was
shown that the scaling~(\ref{kz_zcaling1}) can be derived from the scaling of the
adiabatic fidelity, defined as the overlap of the time-dependent
wave function with the instantaneous ground state:
\be
F(t)=|\langle\psi(t)|\psi_{\rm gs}(t)\rangle|.
\ee
In particular, for $\lambda(t)=\lambda_c+\upsilon t^r/r!\theta(t)$, where $\theta(t)$
is a step function, 
\be P_{\rm ex}(\upsilon)=1-F(t)^2\approx L^d
|\upsilon|^2 \chi_{2r+2}(\lambda_c), \label{p_ex_1} \ee where \be
\chi_{2r+2}(\lambda)={1\over L^d}\sum_{n\neq 0} {|\langle
n|V|0\rangle|^2\over (E_n-E_0)^{2r+2}}, \label{chi} 
\ee 
is the adiabatic (fidelity) susceptibility of the order $2r+2$ ($\chi_2$ is
the conventional fidelity susceptibility~\cite{gu_09}). Here $E_n$
are the eigenenergies  and $V$ is the operator coupled to the
parameter $\lambda$: $V=\partial_\lambda
H(\lambda)|_{\lambda=\lambda_c}$. If the perturbation is local and
spatially uniform , i.e. $V=\int d^d x u(x)$, then the scaling
dimension of the adiabatic fidelity susceptibility is obtained from
a straightforward generalization of the result of
Refs.~\cite{venuti_2007, gu_09}, i.e.
$\dim[\chi_{2r+2}]=2\Delta_u-2z(r+1)-d$, where $\Delta_u\equiv
\dim[u]$ is the scaling dimension of $u(x)$.

Let us now discuss from this general perspective the arguments
leading to the generalized scaling relation Eq.~(\ref{kz_zcaling1}).
If the scaling dimension of the susceptibility is negative, this
implies that $\chi_{2r+2}$ diverges at the critical point. In this
case from Eq.~(\ref{p_ex_1}) we find that asymptotically at
$\upsilon\to 0$ \be P_{\rm ex}(\upsilon)\sim |\upsilon|^2
L^{2d+2z(r+1)-2\Delta_u}. \label{p_ex_2} \ee From Eq.~(\ref{p_ex_2})
we see that the probability of exciting the system becomes of
order one when $L\sim 1/|\upsilon|^{1/(d+z(r+1)-\Delta_u)}$. This
length scale can be interpreted as the typical distance between
elementary excitations (defects) and thus we find that instead of
Eq.~(\ref{p_ex_2}) we get \be n_{\rm ex}\sim
|\upsilon|^{d/(d+z(r+1)-\Delta_u)}. \label{n_ex_2} \ee 
This expression reduces to Eq.~(\ref{kz_zcaling1}) if $u(x)$ is a relevant operator driving the system to the new phase. Indeed, in this case $\lambda
\int d^dx u(x)$ should have the same scaling dimension as the gap,
i.e. $z$, which immediately implies that the scaling dimension of
$u(x)$ is $\Delta_u=d+z-1/\nu$ and that
$\dim[\chi_{2r+2}]=d-2zr-2/\nu$~\cite{degrandi_2010,
schwandt_alet_2009}. Notice finally that if the scaling dimension of
$\chi_{2r+2}$ is positive then the asymptotics in Eq.~(\ref{p_ex_2})
gives a subleading correction to the regular analytic part, $P_{\rm
ex}\approx {\rm const}\;L^d\;\upsilon^2$, coming from the high
energy (ultra-violet) contribution to the susceptibility. We will
discuss its importance in the next section.

Other possible generalizations of the scaling law
Eq.(\ref{kz_zcaling}) involve studies of defect production in
systems where the dynamics describes the passage through quantum
critical lines. A concrete example of such a situation occurs in the
transverse-field XY model \cite{mukherjee_2008, divakaran_2008}.
Here the quench takes one through a gapless line where the critical
point occurs at the same momenta ($\vec k=0$ for the present case)
at each point on the line. A detailed analysis shows that in such
cases, for critical lines with $z=\nu=1$, the defect density
still obeys a universal scaling law albeit with a different power: $n
\sim \upsilon^{1/3}$ \cite{mukherjee_2008, divakaran_2008,
mondal_2009}. The second situation involves the 2D Kitaev model
~\cite{sengupta_2008}, where a quench once again
involves the passage through a gapless line with an energy gap
vanishing for different momenta at different points on the line. It
can be shown that in such a case, the defect density scales as
$\sqrt{\upsilon}$ for 2D Kitaev model instead of the expected $n
\sim \upsilon$ behavior for 2D systems with
$z=\nu=1$~\cite{sengupta_2008}. A generalization of these results
for linear quenches through critical lines with arbitrary $z$ and
$\nu$ has also been worked out \cite{mukherjee_2008, mondal_2008,
mondal_2009, sengupta_2008}. Many other situations involving
anisotropic phase transitions and quenching through multi-critical
points were analyzed in literature leading to various deviations
from the scaling~(\ref{kz_zcaling1})~\cite{deng_2008, bermudez_09,
sen_10}.

In order to detect experimentally the density of excitations
generated by passing through a QCP, it is evident that one should
distinguish between situations where such excitations are long-lived
quasi-particles (as for nearly integrable systems) or decay after being
created (as for non-integrable systems). In the first case, the
presence of excitations above the ground state could for example be
detected by measuring correlation functions long after the quench.
This has been shown for a Quantum Ising chain linearly tuned through
its quantum critical point~\cite{cherng_06}. The presence of defects
with respect to the ferromagnetic ground state lead to exponentially
decaying correlations of the order parameter superimposed, for slow
enough quenches, to characteristic oscillations with period scaling
with the quench velocity. This second feature is
observed for abrupt quenches as well~\cite{sengupta_04} and is a
consequence of the integrability of the model~\cite{rossini_10}. If
in turn we consider a generic non-integrable system, it is necessary
to express deviations from adiabaticity in terms of quantities, such
as the excess energy or the entropy generated by passing through the
QCP, which are not sensitive to the decay of quasi-particles, but
can still be related to the density of excitations created close to
the quantum critical point. Energy can be unambiguously determined
both experimentally and numerically for both integrable and
non-integrable systems and its scaling with the rate of quench can
be used to differentiate between different nonadiabatic
regimes~\cite{polkovnikov_gritsev_08, eckstein_10,moeckel_10}. The
excess energy or equivalently heat ($Q$)~\cite{ap_heat} generated
during the quench process is in general universal if the process
ends near the critical point. The scaling of $Q$ is associated with
the singularity of the susceptibility $\chi_{2r+1}$ at the critical
point ~\cite{degrandi_2009}, which implies that for relevant
perturbations in the thermodynamic limit \be Q\sim
|\upsilon|^{(d+z)\nu/(z\nu r+1)}. \label{heat_scaling} \ee Unless
one considers cyclic processes, the drawback of using heat as a
measure of non-adiabaticity is that it is hard to separate it from
the adiabatic part of the energy change, corresponding to the limit
$\upsilon\to 0$. Moreover, if the position of the QCP is not exactly
known, the heat becomes sensitive to the nonuniversal details of the
spectrum at the final point of the evolution. A way out could be to
measure the higher moments of the energy or the whole distribution function of the energy, connected to
the statistics of the work~\cite{talkner_07, silva_08} in finite size systems. In particular, in the case of abrupt
quenches close to critical points the statics of the work is
characterized by sharp edge singularities~\cite{silva_08,silva1}. A
related natural measure of non-adiabaticity is obtained by focusing
on the entropy since entropy is conserved only for slow (adiabatic)
processes, while  is expected to increase as the system passes
through the QCP. Moreover entropy production can be detected
experimentally in certain systems, e.g. in cold atoms by driving the
system to the weakly interacting regime, where the relation between
entropy and energy is known~\cite{luo_07}. Theoretically, the
quantification of entropy production in a closed quantum system is
rather subtle. Indeed, the von Neumann's entropy of the entire
system, being conserved throughout unitary evolution~\cite{LL5},
cannot be a good characterization of deviations from adiabatic
dynamics. However, the concept of diagonal-entropy~\cite{ap_entropy}, defined as $S_d=-\sum_n
\rho_{nn}\ln(\rho_{nn})$, where $\rho_{nn}$ are the diagonal matrix
elements of the density matrix in the instantaneous basis, avoids this difficulty. In
stationary systems, the diagonal entropy is nothing but the von
Neumann's entropy of the time averaged density matrix, also called
diagonal ensemble. It is clear that the diagonal entropy is
generated only due to nonadiabatic transitions and thus satisfies
the key requirement of the thermodynamic entropy: it is conserved for adiabatic processes, and can only increase or
stay constant in closed systems if the initial state is stationary~\cite{ap_entropy}. For initial equilibrium states the diagonal entropy also satisfies fundamental thermodynamic relation: $dE=TdS-Fd\lambda$, where $F=-\langle \partial_\lambda H\rangle$ is the generalized force. For particular noninteracting models, the
scaling of the diagonal entropy was found to be the same as that of the density of quasi-particles~(\ref{kz_zcaling1})
~\cite{degrandi_10a, degrandi_2010, mukherjee_2008}.
It is also possible to analyze the entanglement entropy~\cite{vidal_03a, refael_04, calabrese_04, calabrese_05}, i.e.  the von Neumann's entropy of the reduced density matrix of a part of the
system, and in particular at its time evolution following a quench~\cite{cincio_07, pollmann_10, sengupta_09}.
For specific 1D spin systems it was found that the entanglement entropy scales logarithmically with the quench
time~\cite{cincio_07, pollmann_10}. Notice however that, at the moment, it is unclear how one can measure entanglement in many body systems and the entanglement entropy in particular (see some suggestions in Refs.~\cite{klich_06, klich_09}) and what its relation with the thermodynamic entropy is.

Finally, another interesting question that has received attention is the connection between microscopic dynamics and
thermodynamics in the semiclassical limit. In general, in classical
systems there is no simple analogue to the instantaneous energy
levels, the key concept in analysis of quantum systems. Such
analogue, however, does exist in the case of periodic motion. Then
in the semiclassical limit the stationary levels are found from the Bohr
quantization (or more accurately from the WKB approximation~\cite{LL3}),
which states that the reduced action in the stationary orbit should
be quantized. In classical mechanics it is known that the reduced
action is an adiabatic invariant, i.e. it is conserved for the
adiabatic evolution~\cite{LLI}. From the previous discussion applied
to quantum systems we can deduce that near singularities like second
order phase transitions, conservation of adiabatic invariants should
be violated and this is indeed the case~\cite{LLI}. In
Refs.~\cite{altland_09, itin_09, itin_09a} slow dynamics was
analyzed for a particular many-body generalization of the
Landau-Zener model (closely related to the Dicke model) in the
semiclassical limit. It was found that the nontrivial power
law scaling of the number of excitations in this system (similar to Eq.~(\ref{kz_zcaling})) follows from the changes of
adiabatic invariants near the singularity, which in turn corresponds to a
quantum critical point in the thermodynamic limit. It is
interesting that quantum fluctuations in this problem entered only
through the initial distribution of the adiabatic invariants but not
through the equations of motion. The corresponding truncated Wigner approximation turned out to be very
accurate in all regimes of the dynamics~\cite{altland_09, kiegel_thesis}. It would be very important to understand precise connections between transitions among microscopic energy levels in the quantum case and changes of suitable generalizations of adiabatic invariants in the classical limit.



\subsection{Slow dynamics in gapped and gapless systems.}

Up to now we have discussed the universal dynamics resulting from
the variation of a control parameter $\lambda$ through a quantum
critical point. However the dynamics of interacting quantum systems
has interesting regimes even when the system is fully gapped or
gapless for the entire duration of the protocol. The
classification of these regimes is important in order to understand dissipation and
to develop optimal protocols minimizing non-adiabatic effects. Many of these
questions are currently a subject of intense theoretical research in
different contexts, from quantum computation to transport.

The general formulas Eq.(\ref{p_ex_1})-(\ref{chi}), which describe
the density of excitations $P_{ex}(\upsilon)$ generated by a
variation of the control parameter, tell us that  if the system
remains fully gapped throughout the evolution, then $P_{\rm ex}$ and
$n_{\rm ex}$ will have a quadratic scaling with $\upsilon$ whenever
the susceptibility $\chi_{2r+2}$ evaluated at the initial and final
couplings is finite~\cite{degrandi_2009, ortiz_2008}. A similar
argument shows that the heat $Q$ is also quadratic in $\upsilon$ if
$\chi_{2r+1}$ is finite. This quadratic scaling is characteristic of
any quantum system. Let us point that in the standard Landau-Zener problem in the slow limit $\Delta^2\gg \upsilon$ if we start in the ground state at $t \to -\infty$ and let the system  evolve up $t \to +\infty$, the  transitions to excited states are exponentially suppressed as a result of destructive interference between multiple  transitions~\cite{vitanov_96, vitanov_99}. At the same time in the intermediate stages of the evolution the transition probability reaches much higher values which scale only quadratically with the rate $\upsilon$. For example, if one
considers a process which starts at $t_0\to-\infty$ the transition probability to the instantaneous excited state at the moment $t$ in the slow limit $\upsilon/\Delta^2\ll 1$ can be approximated by~\cite{vitanov_99}:
\be
p_{\rm ex}\approx {\upsilon ^2
\Delta^2\over 16 (g(t)^2+\Delta^2)^3}. \label{lz_prob_1}
\ee
If $t>0$ there is an additional exponential term which leads to
Eq.~(\ref{zl_prob}) at $t\to+\infty$. If the process starts at $t_0=0$ exactly
in the symmetric point, where $g(0)=0$ the quadratic asymptotics~(\ref{lz_prob_1}) is also recovered
~\cite{cucchietti_07}. This scaling occurs as a result of the
discontinuity of the first derivative of $g(t)$ at the moment where
the process starts or ends or following a discontinuity in
any other point of the evolution (see e.g. Refs.~\cite{damski_06,
divakaran_10} for particular cases). Likewise if there is a
discontinuity in the second order derivative of $g(t)$
asymptotically the transition probability in the LZ problem scales
quadratically with acceleration. More generally for the protocol
$g(t)=g_0+\upsilon (t-t_0)^r/r!\;\theta(t-t_0)$, where
$\theta(t)$ is the step function, one can show
that~\cite{degrandi_2009}
\be
p_{\rm ex}(t\to\infty)\approx
{\upsilon^2\Delta^2\over 16 (g_0^2+\Delta^2)^{2r+1}}.
\label{lz_prob_2}
\ee
As we discussed in the previous section this formula applies even to sudden transitions ($r=0$) where it reduces
to the result of the ordinary perturbation theory. The same expression applies to the reverse process. If both the initial and final couplings are finite then the resulting transition probability is asymptotically determined by the sum of probabilities associated with discontinuities of derivatives of $g(t)$ at the initial and
final times of the evolution plus additional interference terms
which highly oscillate in the slow limit. Let us point out that in the
LZ problem (and in general in gapped systems) one can suppress power
law asymptotics of the transition probability by starting and ending
the protocol infinitely smoothly, e.g. $g(t)\sim
g_0+g_1\exp[-\tau/(t-t_0)]\theta(t-t_0)$. In this case only the
non-analytic term in the transition probability survives and we are
back to Eq.~(\ref{zl_prob}) where $\upsilon$ is the time derivative of
$g(t)$ near the symmetric point where $g(t)=0$.

In gapless systems the situation becomes qualitatively different. In this case the adiabatic
susceptibilities can diverge leading to non-analytic dependence of the corresponding quantities on $\upsilon$, as
in the case of the crossing of a quantum critical point. For example, it is straightforward to see that for marginal
perturbations in a generic gapless phase the scaling dimension of the adiabatic susceptibility $\dim[\chi_{2r+2}]=d-2zr$. It becomes negative in low dimensions $d<2zr$ leading to a non-analytic
scaling of the density of excitations with $\upsilon$. Thus depending on dimensionality in gapless systems one expects at least two different regimes of the response of the system to a slow external perturbation: analytic and non-analytic. These regimes were first suggested in Ref.~\cite{polkovnikov_gritsev_08} together with a third regime where
adiabaticity is violated in the thermodynamic limit and $Q$ or $n_{\rm ex}$ become proportional to a power of the system size or some other large length scale associated with some irrelevant operator. In this regime,  which can be realized
in low-dimensional bosonic systems, the scaling Eq.~(\ref{kz_zcaling}) is violated. At the moment it is unclear how
generic it is and what sets the scaling of various quantities.


A close inspection of the adiabatic susceptibility shows that in general the analytic (quadratic)
part of the heat and energy of excitations on $\upsilon$ comes from the high energy (or ultra-violet) part of the spectrum, while the non-analytic part comes from low energies. This was indeed shown to be the case in several situations, from the sine-Gordon model~\cite{degrandi_2010}, to the Falicov-Kimball model~\cite{eckstein_10}
and the turning on interactions in a Fermi liquid~\cite{moeckel_10}. As we pointed above the ultra-violet transitions can be suppressed by avoiding discontinuities in $\lambda(t)$ and its derivatives. However, this is not necessarily the case for the low energy non-analytic contribution. To see this we need to reexpress  the excess energy (or density of excitations) in terms of the total time of the process $\tau$. Doing this  it was found that in an insulating, gapped phase, the details of the protocol are important and  smoother protocols lead to a suppression of
non-adiabatic effects, while in a gapless phase making $\lambda(t)$ smoother does not affect the heating~\cite{eckstein_10}. This result can be again understood by analyzing the scaling dimension of the
susceptibility $\chi_{2r+1}$. According to our discussion for generic gapless systems its scaling dimension is negative when $d+z<2zr$. Then $Q\sim \upsilon^{(d+z)/zr}$ (this result immediately follows from Eq.~(\ref{heat_scaling}) by
taking the limit $\nu\to\infty$). On the other hand, $\upsilon$ is related to the total quench time as $\upsilon\sim 1/\tau^r$. Thus we see that in this case $Q\sim 1/\tau^{(d+z)/z}$, i.e. indeed independent on $r$. On the other hand for positive scaling dimension of $\chi_{2r+1}$ which is the case for $d+z>2zr$ and which is always true in gapped systems we have $Q\sim \upsilon^2\sim 1/(\tau)^{2r}$. Since in the adiabatic limit $\tau$ is large we see that indeed the heat can be suppressed by increasing $r$ and making $\lambda(t)$ smoother. An interesting open question is finding an optimal protocol for minimizing the non-adiabatic effects within given time $\tau$. It is plausible that the optimal power is determined by a vanishing scaling dimension of the corresponding adiabatic susceptibility $\chi_{2r+1}$.
The questions of finding protocols minimizing non-adiabatic effects for gapped systems (with possibility of
crossing isolated quantum critical points) were also addressed  by approximately minimizing the transition
probability  and identifying the Riemannian metric tensor underlying the adiabatic evolution~\cite{rezakhani_09, rezakhani_10}. Studying the optimization of the protocol taking a system through a QCP it was found the optimal exponent of $\lambda(t)\sim |t|^r{\rm sign}(t)$ near the QCP scales logarithmically with the quench time, $r\propto \ln(\tau)$~\cite{optimal_passage}. This result was also extended to systems with external confining potential~\cite{collura_10} .

\subsection{Effects of finite temperature.}

In the discussion above we always implicitly assumed that the system is initially prepared in the ground state. An interesting and genuine question is how finite temperature effects modify the picture.
In isolated systems temperature enters through initial conditions: the system is prepared
in the initial finite temperature equilibrium state  and is then dynamically driven out of equilibrium.
How is the response of the system affected by the initial thermal fluctuations ? One naturally expects that
while the transitions to high energy states (quadratic in $\upsilon$) will not be affected by small temperatures in the system, the transitions to the low energy states, which determine the non-analytic contribution to heat and density of excitations, will be very sensitive to temperature. In
Refs.~\cite{degrandi_10a, degrandi_2010} (see also Ref.~\cite{gritsev_10}) studying a particular Sine-Gordon model
in the two limits where it could be mapped to free bosons and free fermions, it was shown that the statistics of
quasi-particles enters the scaling of both $Q$ and $n_{\rm ex}$ making dynamics more adiabatic (compared to the zero temperature case) for fermions due to Pauli blocking and less adiabatic for bosons due to Bose enhancement. These results were not yet extended to generic interacting systems.

Another aspect of thermalization, the influence of the coupling to an environment setting the temperature on the
slow dynamics near quantum critical points, has been studied in Refs.~\cite{patane_08, patane_09, patane_09a, mostame_07,fubini_07}. This setup allows one to analyze the effects of thermal smearing and of dephasing/dissipation on the dynamics of a quantum critical system. Using a combination
of kinetic equations and scaling arguments it was found that in this situation the excess energy has two universal  contributions, one still given by Eq.~(\ref{heat_scaling}), while the second involving a universal power
of temperature replacing the universal power of $\upsilon$~\cite{patane_08}.

\subsection{Open problems}

While the physics described above is definitely an important example of the emergence
of universality in the dynamics of interacting quantum systems, it is evidently
a piece, albeit important, of the puzzle that has to be composed in order to
understand to which extent the standard concepts of statistical physics can be applied
to nonequilibrium problems. Understanding the meaning of  \it relevance or irrelevance \rm of a perturbation in generic nonequilibrium processes,  extending the notion of universality to nonequilibrium systems, as well as the concept of
renormalization group, is a task (or dream) that certainly requires the solution of many specific problems, and
a close comparison between experiments and theory.

So far most of the theoretical research focused on analyzing slow dynamics for global quenches, where the external perturbation couples to the whole system. How these results can be extended to local or spatially nonuniform perturbations is an open question.  At one extreme limit, one can imagine performing a quench only
locally. Then the rest of the system could be seen as a thermal bath. The analysis of a special case of dynamics of a transverse field Ising model where the tuning parameter linearly depends both on time and space has shown that
excitations are generally suppressed by nonuniformity of the tuning parameter~\cite{dziarmaga_10a}. This suggests that quantitative and qualitative differences may emerge when some of the symmetries of the system, e.g. translational, are broken in the quench process.

Another important issue concerns the
connections between adiabaticity in thermodynamics and microscopic dynamics. One of the consequences of the thermodynamic adiabatic theorem is that no heat can be generated in an isolated system during an infinitesimally slow process. More generally according to the second law of thermodynamics in the Thompson's (Kelvin's) form for any cyclic process the system can only increase its energy, i.e. the heat should be always non-negative as long as one starts in equilibrium. This statement, which is obvious if the system is initially in the ground state, has been proven microscopically for a class of passive initial states~\cite{thirring,allahverdyan_02, allahverdyan_05, boksenbojm_09}, whose initial density matrix is stationary (diagonal) and monotonically decreasing function of energy: $(\rho_n-\rho_m)(\epsilon_m-\epsilon_n)\geq 0$.
This statement also directly follows from analyzing transitions between microscopic energy levels~\cite{ap_heat}.
Likewise many statements of thermodynamics related to behavior of entropy  including the second law and fundamental thermodynamic relations are recovered using the conecpt of diagonal entropy~\cite{ap_entropy}. At the same time there are many open questions remaining: what are the time scales involved in the definition of adiabaticity ? How one can microscopically define adiabatic time scales in interacting systems and why these time scales are much shorter than inverse distance between many-body levels (see e.g. discussion in Ref.~\cite{balian}) ?  And finally, what is the role of integrability in nonequilibrium thermodynamics? These questions are closely connected to the microscopic origin of conventional dissipation, which in turn is also very likely related to the combination of nonadiabatic creation of the elementary excitations and their following relaxation or dephasing. From the discussion above, we can anticipate anomalous dissipation near critical points and in gapless low-dimensional systems.

\section{Effects of Integrability and its breaking: ergodicity and thermalization.}
\label{secIII}

Let us now turn to one of the most natural questions to be addressed when studying the
dynamics of a closed many-body quantum system: are
interactions within the system sufficient to make the system
behave ergodically ? If we focus on local degrees of freedom, e.g.
a few spins in a spin chain, can the rest of the system be always thought as an effective thermal bath? And if this is not possible, are there some observable  effects on the system dynamics?
While these questions are definitely connected to quantum
ergodicity~\cite{goldstein_10}, a topic with a long history dating
back to the early days of quantum
mechanics~\cite{neumann_29,pauli_37,mazur_68,suzuki_70,peres_84,deutsch_91,srednicki_94},
the past few years have brought a great deal of progress in the
context of closed many-body systems. The main motivation came from
recent experiments on low dimensional cold atomic gases described in
some detail in Sec.~\ref{sec:exp} in this Colloquium~\cite{greiner2002b,kinoshita}. The experimental
availability of essentially closed (on the time scales of experiments) strongly correlated systems
together with the awareness of the conceptual importance of these issues in a number of areas (e.g. transport problems, many-body localization, integrable and non-integrable dynamics) have stimulated a lot of interest on \it quantum thermalization\rm. Below we will give a synthetic view on a number of recent important developments on this subject, starting with the discussions of the general concepts of ergodicity and thermalization, and then moving to the discussion of many-body systems and integrability.

\subsection{Quantum and classical ergodicity.}

While the idea of ergodicity is well defined in classical mechanics,
the concept of \it quantum ergodicity \rm is somewhat less precise
and intuitive. Classically, an interacting system of $N$ particles
in $d$ dimensions is described by a point $X$ in a
$(2\;d\;N)$-dimensional phase space. The intuitive content of the
word "ergodic", i.e. the equivalence of phase space and time
averages, can be then formalized by requiring that if we select an
initial condition $X_0$ having initial energy $H(X_0)=E$, where $H$
is the Hamiltonian of the system, then
\begin{equation}\label{micro_classical}
\overline{\delta(X-X(t))}\equiv \displaystyle\lim_{T\to\infty}\frac{1}{T}\int_0^{T}dt\,\delta(X-X(t))=\rho_{\rm mc}(E),
\end{equation}
where $\rho_{\rm mc}(E)$ is the microcanonical density of the system on the hyper-surface
of the phase space of constant energy $E$, and $X(t)$ is the phase space trajectory with initial condition
$X_0$. Of course if this condition is satisfied by all  trajectories, then it is also true for every observable. We immediately see that in order to have ergodicity, the dynamics cannot be arbitrary: the trajectories $X(t)$ have to cover uniformly the energy hyper-surface for almost every initial condition $X_0$.

The most obvious quantum generalization of this notion of ergodicity
is arduous~\cite{neumann_29}. Let us first of all define a quantum
microcanonical density matrix: given a Hamiltonian with eigenstates
$\mid \Psi_{\alpha} \rangle$ of energies  $E_{\alpha}$, a viable
definition of the microcanonical ensemble is obtained by coarse graining the spectrum on energy
shells of width $\delta E$, sufficiently big to contain many states
but small on macroscopic scales. Denoting by ${\cal H}(E)$ the set of
eigenstates of $H$ having energies between $E$ and $E+\delta E$, we define $\hat{\rho}_{\rm mc}(E)=\sum_{\alpha \in {\cal H}(E)} 1/{\cal
N}\mid \Psi_{\alpha} \rangle \langle \Psi_{\alpha} \mid$, where
${\cal N}$ is the total number of states in the micro-canonical shell. Let us now ask the most obvious question: given a generic initial condition made out of states in a microcanonical shell,
$\mid \Psi_0 \rangle=\sum_{\alpha \in {\cal H}(E)}\; c_{\alpha}\;
\mid \Psi_{\alpha} \rangle$, is the long time average of the density
matrix of the system given by the microcanonical density matrix? The
answer to this question for a quantum system is, unlike in the
classical case, almost always no, as J. von Neumann realized already
in 1929~\cite{neumann_29}. More precisely, if we assume the
eigenstates of the system not to be degenerate,  the time average is
\begin{eqnarray}
\overline{\mid \Psi(t) \rangle \langle \Psi(t) \mid}=\sum_{\alpha}
\mid c_{\alpha} \mid^2 \mid \Psi_{\alpha} \rangle
\langle \Psi_{\alpha} \mid=\hat{\rho}_{\rm diag},
\end{eqnarray}
where $\mid \Psi(t) \rangle$ is the time evolved of $\mid \Psi_0
\rangle$. This object is known in the modern literature as the \it
diagonal ensemble \rm~\cite{rigol_07,rigol_08,rigol_09}. Notice now
that the most obvious definition of ergodicity, i.e. the requirement
$\rho_{\rm mc}=\rho_{\rm diag}$, implies that $\mid c_{\alpha}
\mid^2=1/{\cal N}$ for every $\alpha$, a condition that can be
satisfied only for a very special class of states. Quantum
ergodicity in the strict sense above is therefore almost never
realizable~\cite{neumann_29,goldstein_10}.

Our common sense and expectations, which very frequently fail
miserably in the quantum realm, make us nevertheless believe that,
in contrast with the arguments above, macroscopic many-body systems
should behave ergodically almost always, unless some very special
conditions are met (e.g. integrability). The key to understand
ergodicity in therefore to look at quantum systems in a different way,
shifting the focus on {\it observables} rather than on the states
themselves~\cite{neumann_29,peres_84,mazur_68}. Given a set of \it
macroscopic observables \rm $\{M_{\beta}\}$ a natural expectation
from an ergodic system would be for every $\mid \Psi_0 \rangle$ on a
microcanonical shell ${\cal H}(E)$
\begin{eqnarray}\label{req1}
\langle \Psi(t) \mid M_{\beta}(t)\mid \Psi(t) \rangle \displaystyle
\rightarrow_{t\rightarrow +\infty} {\rm Tr}[M_{\beta} \hat{\rho}_{\rm mc} ]
\equiv  \langle M_{\beta} \rangle_{\rm mc},
\end{eqnarray}
i.e. that looking at macroscopic observables long after the time
evolution started makes the system appear ergodic for every initial
condition we may choose in ${\cal H}(E)$. One needs a certain care in defining the infinite time limit here, since literally speaking it does not exist in finite systems because of quantum revivals. A proper way to understand this limit is to require that Eq.~(\ref{req1}) holds in the long time limit at almost all times. Mathematically this means that the mean square difference between the LHS and RHS of Eq.~(\ref{req1}) averaged over long times is vanishingly small for large systems~\cite{reimann_08}. To avoid dealing with these issues ergodicity can be defined using the time average, i.e. requiring that
\begin{eqnarray}\label{req2}
\overline{\langle \Psi(t) \mid M_{\beta}(t)\mid \Psi(t)\rangle} = {\rm Tr}[M_{\beta} \hat{\rho}_{\rm diag} ]
= \langle M_{\beta} \rangle_{\rm mc}.
\end{eqnarray}
Notice that if the expectation value of $M_{\beta}$ relaxes to a well defined state in the sense described above, this state will coincide with the time averaged state and the two definitions of ergodicity Eq.(\ref{req1})-(\ref{req2}) will be equivalent. If the conditions above are satisfied then in loose terms $\hat{\rho}_{\rm mc}$ can be considered as equivalent to $\hat{\rho}_{\rm diag}$\,.  J. von Neumann proved that if the system
satisfies some very natural requirements (e.g. absence of
resonances), and the set $\{M_{\beta}\}$ is constructed in such a
way as to define \it macrostates \rm of the system, which obviously
requires  the observables to be coarse grained on the various
microcanonical shells ${\cal H}(E)$ and mutually commuting, then a
form of ergodicity is observed (sometimes referred to a {\it normal
typicality}). In particular, for every $\mid \Psi_0 \rangle$ and
\it almost \rm every set $\{M_{\beta}\}$ the diagonal and
microcanonical ensembles are equivalent~\cite{goldstein_10,neumann_29}. More recently it was proven that the whole density matrix of a small subsystem of a bigger system which is placed initially in a typical eigenstate is described by the canonical ensemble~\cite{popescu_10}. In Ref.~\cite{gogolin_11} these results were further extended to the problem of measurement and decoherence. Particular care is nevertheless needed in relating  these statements to the dynamics and thermalization of actual many body systems, since physical initial conditions in quenched system almost never correspond to eigenstates of a new Hamiltonian.

\subsection{Nonergodic behavior in integrable systems: the generalized Gibbs ensemble}

While the statements above are very general, their application to
specific systems is not at all straightforward. Looking at a
concrete many-body system, it is of primary interest not just to
find out whether in principle a set of macroscopic observables that
behave ergodically exists, but whether specific and natural observables, such as the magnetization for spin
chains, density for cold atomic gases, or various correlation functions behave ergodically or not. In this respect, experiments tell us that ergodicity is not at all guaranteed~\cite{kinoshita} if the closed system is integrable or nearly integrable. While this fact was expected~\cite{mazur_68,suzuki_70,girardeau_69,girardeau_70, barouch} recent research on the dynamics of integrable systems has focused on finding ways to predict the asymptotical states taking into account integrability, i.e. the presence of many constants of motion.



Let us discuss how this can be done qualitatively using the simplest
example of integrable system, a periodic harmonic chain of finite
length described by the Hamiltonian
\be
H=\sum_{j=1}^{M-1} \left[{p_j^2\over 2m}+{m\nu^2\over 2}(x_j-x_{j+1})^2\right],
\ee
where $x_j$ are deviations of particles from the equilibrium
positions and $p_j$ are their momenta; we use the identification
$x_M\equiv x_0$. Let us imagine that initially we deform the system
in a particular way and ask how this deformation evolves with time.
We note that since this is a harmonic system described by linear equations of motion the following analysis also applies to quantum systems. From elementary physics we know that the initial deformation splits into normal modes characterized by the quasi-momenta $q_n=2\pi n/M$, where $n$ is integer $n\in[0,M-1]$, and the dispersion $\omega_q=2\nu\sin{q/2}$. This system obviously does not thermalize even at long times because there is no energy exchange between modes. This does not imply though that it can not reach a well defined asymptotic  state in the long time limit~\cite{barthel_08, cramer_10}. To illustrate this point consider for example the displacement of the $j$-th atom at time $t$ after some initial displacement:
\be
x_j(t)={1\over \sqrt{M}}\sum_{n=0}^{M-1} x_{q_n}(t) \mathrm e^{i q_n j}.
\label{x_j}
\ee
$x_q(t)=A_q\cos[\omega_q t]$, where $A_q$ is a complex amplitude determined by initial conditions (for simplicity we assumed an initial stationary state).  Let us now analyze qualitatively the dynamics of this system. At short times, provided that the initial modulation is smooth and only modes with small momenta ($q\ll \pi$) are excited we can linearize the spectrum $\omega_q\approx \nu q$. Clearly in this case we recover periodic motion of the wave-packet with a period equal to the ratio of the system size and sound velocity: $T=M/\nu$. This persistent motion is characteristic of the absence of any relaxation. However, as time gets longer deviations of the dispersion from linear become more important. In particular, when $t^\ast(\omega_{\overline{n}+1}+\omega_{\overline{n}-1}-2\omega_{\overline{n}})\sim
1$, where $\omega_{\overline{n}}$ is the central frequency of the wavepacket, correlations between phases among the different modes are lost and they can be treated as essentially random numbers. For our model $t^\ast\sim M^2/\omega_{\overline{n}}$. At long times $t\gg t^\ast$ the different momentum modes become uncorrelated and the system reaches the asymptotic stationary state in a sense we defined earlier (it can be found close to that state at almost all times). For this asymptotic state the only relevant information about the initial conditions is encoded in the $M$
mode amplitudes $|A_q|$ or equivalently in their squares $|A_q|^2$ proportional to the occupancies of the modes of energy $E_q$, which are the integrals of motion. Note though that there are special modes corresponding to momenta $q$ and $-q$ which are exactly degenerate. The correlations between $A_q$ and $A_{-q}=A_q^\star$ thus never disappear and in general one needs to fix $M$ additional constraints representing the relative phases of the complex amplitudes. For example, if the initial configuration is symmetric $x_j=x_{-j}$ then $A_q$ is real, meaning that the phases of all modes are identical.  Then it is easy to check that with this constraint $\langle x_j^2(t)\rangle$ acquires spatial dependence on $j$ even in the long time limit. This dependence can not be recovered by fixing only mode occupancies. Only when these phases are unimportant e.g. they average to zero or if there are no degeneracies between the normal modes the asymptotic state is fully fixed by the integrals of motion. Thus in contrast with ergodic systems where only the energy needs to be fixed, the long time behavior of our integrable model can be reproduced by fixing the $M$ integrals of motion and possibly $\sim M$ other constraints if there are degeneracies.  While the number of commuting (local) integrals of motion is large, equal to the system size $M$, it is vastly smaller than the total number of states which scales with $M$ exponentially.

Let us now see how these considerations are transposed in
many-body systems, focusing on another simple integrable model,  the
Quantum Ising chain~\cite{Sachdev_book} described by the
Hamiltonian: $H_0=-\sum_i \sigma^{x}_i \sigma^{x}_{i+1}+g
\sigma_i^z$. Here $\sigma^{x,z}_i$ are the spin operators at site
$i$ and $g$ is the strength of the transverse field. This model gives one of the simplest examples of
quantum phase transition, with a quantum critical point at $g_c=1$ separating two mutually dual gapped phases, a quantum paramagnet ($g>g_c$) and a ferromagnet ($g<g_c$).

In the Quantum Ising chain the \it local \rm transverse magnetization, $M^{x}=\sum_i \sigma_i^{x}$ is a non-ergodic
operator~\cite{mazur_68,girardeau_69,girardeau_70,barouch}. To see this, it is useful to employ a Jordan-Wigner transformation that reduces the problem to a free fermion model~\cite{Sachdev_book}. In terms of the
fermionic operators $c_k$ relative to modes of momentum $k=2\pi n/L$ the Hamiltonian takes the form
\begin{eqnarray}
H&=&2\sum_{k>0} (g-\cos(k))(c^{\dagger}_kc_k-c_{-k}c^{\dagger}_{-k})\nonumber \\
&+& i\sin(k)(c^{\dagger}_kc^{\dagger}_{-k}-c_{-k}c_{k}),
\end{eqnarray}
Under this mapping the transverse magnetization becomes
$M^{x}=-2\sum_{k>0} (c^{\dagger}_kc_k-c_{-k}c^{\dagger}_{-k})$. The
eigenmodes $\gamma_k$ of energy $E_k=2\sqrt{(g-\cos(k))^2+\sin(k)^2}$
diagonalizing the Hamiltonian are related to the fermionic operators
$c_k$ by a Bogoliubov rotation,
$c_k=\cos(\theta_k)\gamma_k-i\sin(\theta_k)\gamma^{\dagger}_{-k}$,
with $\tan(2\theta_k)=\sin(k)/(g-\cos(k))$. In Heisenberg
representation the operators $\gamma_k$ acquire simple time
dependence: $\gamma_k(t)=\gamma_k(0)\exp[-i E_k t]$.
As in the previous problem of a harmonic chain if the energies $E_k$ are incommensurate, at sufficiently long times different momentum modes become statistically independent from each other. This statement does not apply to modes with opposite momenta $k$ and $-k$ which have identical energies. However, if these correlations are not important then in the long time limit (see below)  each mode can be characterized by the conserved quantity $n_k=\langle\gamma_k^\dagger\gamma_k\rangle$. Let us now consider unitary dynamics of the transverse magnetization starting with a generic initial condition $\mid \Psi_0 \rangle$. The time evolution of the operator $M^x(t)$ expressed in terms of the eigenmodes of the Hamiltonian is
\begin{eqnarray}
&&M^x(t)=-2\sum_{k>0}\cos(2\theta_k)(\gamma_k^{\dagger}\gamma_k-\gamma_{-k}\gamma^{\dagger}_{-k})
\nonumber \\
&&~~~+ i\sin(2\theta_k)(\gamma_{-k}\gamma_ke^{-2iE_k t} -\gamma^{\dagger}_k\gamma^{\dagger}_{-k}e^{2iE_k t}).
\end{eqnarray}
In the long time limit only the diagonal terms in the sum survive,
while the off diagonal ones, describing creation or destruction of
two fermions average to zero. Therefore for any initial condition
$\mid \Psi_0 \rangle$ the asymptotic value attained by the
transverse magnetization is
\begin{eqnarray}
\overline{ \langle M^x(t) \rangle} =-2\sum_{k>0}\cos(2\theta_k)(\langle \gamma_k^{\dagger}\gamma_k \rangle -\langle \gamma_{-k}\gamma^{\dagger}_{-k} \rangle).
\end{eqnarray}
This asymptotic value is therefore perfectly described by the set of the occupation numbers $n_k$.

The result above leads one to conjecture that the asymptotic state is described by a Gibbs-like statistical ensemble of the type~\cite{rigol_07}
\begin{eqnarray}\label{GibbsIsing}
\rho_{G}=\frac{e^{-\sum_k\lambda_k \gamma^{\dagger}_k\gamma_k}}{Z},
\end{eqnarray}
where the Lagrange multipliers $\lambda_k$ are fixed by requiring that $n_k\equiv\langle \Psi_0 \mid \gamma^{\dagger}_k \gamma_k \mid \Psi_0 \rangle ={\rm Tr}[\rho_{G} \gamma^{\dagger}_k \gamma_k ]=\langle \gamma^{\dagger}_k \gamma_k \rangle_G$.
The ensemble defined above in Eq.~(\ref{GibbsIsing}) can be seen as
a particular case of the ensemble
\begin{eqnarray}\label{GGE}
\hat{\rho}_G=\frac{e^{-\sum_{\alpha} \lambda_{\alpha} I_{\alpha}}}{Z},
\end{eqnarray}
 known as the generalized Gibbs ensemble (GGE) or the maximum entropy ensemble and introduced by Jaynes~\cite{jaynes_57} to describe the equilibrium state of a system possessing $N$ constant of motions ${I}_k$. A recent conjecture~\cite{rigol_07} proposed to use the GGE to describe the asymptotic state of a generic quantum integrable model. This proposal had however to face two obvious subtleties. It is first of all to be specified how to choose
the $I_k$ in Eq.(\ref{GGE}). Indeed, if all constants of motion would be admissible, including non-local ones, then one would obviously and tautologically describe the asymptotic state (both for integrable and non-integrable systems),
as one can easily see  by choosing as $I_k$ the projectors onto the eigenstates of the Hamiltonian.
The way out comes however by observing that in standard thermodynamics the Gibbs ensemble emerges for small subsystems from the assumption of statistical independence between sufficiently big subsystems\rm. In this derivation the additivity of a conserved quantity - energy - plays a crucial role. This is the reason why the probability of a
given configuration is exponential in energy and not e.g. in energy squared ~\cite{kardar}. Similar arguments apply to any additive integrals of motion so that statistical independence and invariance of the ensemble to the choice of a subsystem of an integrable system puts strong constraints on the choice of the integrals of motion in
GGE when the latter is applied to subsystems of an integrable system. In this respect the average occupation numbers of different momentum modes used in Eq.~(\ref{GibbsIsing}) become approximately additive for small subsystems. This approximate additivity of integrals of motion was recently discovered in Ref.~\cite{cassidy_10} for another integrable system of one-dimensional hard-core bosons. In particular, it was noticed that the integrals of motion $I_\alpha$ and the lagrange multipliers $\lambda_\alpha$ in that case can be written as a smooth functions of $\alpha/N$ implying that in large systems the argument of the exponent entering Eq.~(\ref{GGE}) can be written in the extensive (additive) form: $\sum_\alpha \lambda_\alpha I_\alpha\approx L\int_0^1 d\xi, I(\xi)\lambda(\xi)$, where $\xi=\alpha/L$. This suggests that GGE can be defined through a smooth function $\lambda(\xi)$, which replaces the temperature in the ergodic systems.

There is a second subtlety in applying the GGE to quantum systems. Here the most natural definition of integrability is based on requirement that the system has well defined quasi-particles that maintain their identity upon scattering (see Ref.~\cite{caux_10} for a more detailed discussion), i.e. scattering is purely elastic and there is no production of particles or dissipation associated to it~\cite{sutherland_04,mussardo_09}. This notion can be made precise in continuum integrable models, such as the Luttinger liquid or the Sinh-Gordon model, which can emerge as low energy descriptions of other integrable models,  such as the critical XXZ chain and the Lieb-Liniger gas. In these systems it is natural to associate the $\hat{I}_{\alpha}$ to the occupation numbers of the quasi-particle states. More specifically, considering a generic one dimensional relativistically invariant integrable system with say a single species of quasi-particles  of mass $m$, energy $E=m\cosh(\theta)$ and momentum $p=m\sinh(\theta)$ ($\theta$ is the rapidity), the quasi-particles can be described by annihilation operators $\hat{A}(\theta)$ satisfying the algebra $\hat{A}(\theta_i)\hat{A}(\theta_j)= S(\theta_i-\theta_j) \hat{A}(\theta_j)\hat{A}(\theta_i)$, where $S$ is the S-matrix of the two particle scattering. Similar relations are valid for the products of creation, and creation-annihilation operators~\cite{mussardo_09}. Since the Hamiltonian is by definition diagonal in $\hat{A}(\theta)$, $H=\int d\theta E(\theta)A^{\dagger}(\theta)A(\theta)$, and every eigenstate can be written as $\mid \theta_1,\dots,\theta_n \rangle=A^{\dagger}(\theta_1)\cdot \dots \cdot A^{\dagger}(\theta_n) \mid 0\rangle$, with $\theta_1 > \dots >\theta_n$, in this case it is rather natural to postulate the form
\begin{eqnarray}
\hat{\rho}_G=\frac{e^{-\int d\theta \lambda(\theta) A^{\dagger}(\theta)A(\theta)}}{Z},
\end{eqnarray}
for the generalized Gibbs ensemble~\cite{fioretto_09}. This ensemble is a direct generalization of the GGE for the Quantum Ising Model, where $S=-1$. For this general class of integrable systems and a specific class of \it translationally invariant \rm initial states it was indeed shown the long-time limit of the \it average \rm of local operators is well described by this ensemble~\cite{fioretto_09}. Such initial states can be written as
\begin{eqnarray}
\mid \Psi_0 \rangle = {\cal N} e^{-\int d\theta\;K(\theta)\;A^{\dagger}(\theta)A^{\dagger}(-\theta)},
\end{eqnarray}
which in turn are similar to the so-called \it integrable boundary states
\rm in statistical field theory~\cite{ghoshal_94}. Such states naturally emerge in
experimentally relevant systems, for example when studying dephasing in split quasi-$1d$ condensates~\cite{gritsev_07}
or in the Quantum Ising model, when studying a quantum quench from a transverse field $\gamma_i$ to a transverse field
$\gamma_f$~\cite{silva_08}.

A very interesting  idea related to the GGE was suggested by Gurarie~\cite{gurarie_95} to explain the steady state of a driven nearly integrable system. It was shown that the steady state distribution of the wave amplitudes corresponding to different momenta (see Ref.~\cite{zakharov} for details) can be obtained by taking the probability density $\rho\propto \exp[-F]$, where $F$ is a (complex) combination of the approximate integrals of motion found
perturbatively. In terms of this ensemble one recovers the correct power law distribution of the
amplitudes of waves with the momentum and other observables.

Another view towards elucidating the validity of the
generalized Gibbs ensemble has been pursued for special quenches in a 1D
Bose-Hubbard model~\cite{cramer_08} and in integrable systems with
free quasiparticles~\cite{barthel_08}. It was shown that, upon
tracing all degrees of freedom of the system outside a small region
of space and under specific conditions, the local density matrix
tends asymptotically to $\hat{\rho}_{G}$. More recently a series of recent theoretical ~\cite{flesch_08}
and experimental~\cite{trotzky_11} works on the dynamics of Bose-Hubbard models has proven the
relaxation of local observables in these system to a maximum entropy ensemble consistent with the
constraints of the dynamics. A hint towards the generalization of $\hat{\rho}_G$ for Bethe Ansatz integrable systems
was proposed in Ref.~\cite{barthel_08}. The GGE was also tested in a number of models, from Luttinger liquids~\cite{cazalilla_quench_06, iucci_09} and free bosonic theories~\cite{calabrese_07}, to integrable hard-core boson models~\cite{rigol_07} and Hubbard-like models~\cite{eckstein_08, kollar_08}. In all cases, it was shown to correctly predict the asymptotic momentum distribution functions for a variety of systems
and quantum quenches.

At this point is should be stressed that as we discussed before the GGE does not always give complete description of the asymptotic state of the system. In the simple example of the harmonic chain we saw that for generic initial conditions it is necessary to specify $2N$ real constants or $N$ complex amplitudes in order to correctly describe the asymptotic state even if we focus exclusively on local observables. For a quantum Ising chain, moreover, $\rho_{G}$ can be interpreted as a grand-canonical distribution with an energy dependent chemical potential $\mu_{k}=E_k-\lambda_k$. It is evident now that if we consider the correlations of $\delta n_k = \gamma^{\dagger}_k \gamma_k - \langle \gamma^{\dagger}_k \gamma_k \rangle$, the occupation numbers of different eigenmodes, the GGE necessarily predicts $\langle \delta n_k \delta n_{k'} \rangle=0$. Likewise the GGE predicts the correlators of the type $\langle\gamma_k^\dagger(t)\gamma_{-k}(t)\rangle$ are always equal to zero. For a generic initial state $\mid \Psi_0 \rangle$ both statements are not necessarily true: by e.g. breaking translational invariance in the initial state one could have $\langle \Psi_0 \mid \delta n_k \delta n_{k'} \mid \Psi_0 \rangle \neq 0$ and $\langle \Psi_0 \mid \gamma_k^\dagger \gamma_{-k} \mid \Psi_0 \rangle\neq 0$. Notice that the mere survival of off-diagonal correlations of this type when the evolution starts with a non-translationally invariant state signals in a sense the integrability of the model, i.e. the existence of well defined quasiparticles $\gamma_k$. Indeed, following the argument of Gangardt and Pustilnik \cite{gangardt1}, if the Hamiltonian of the system is translationally invariant but integrability is broken, the off-diagonal correlators are expected to decay to zero for any initial condition, thereby restoring the translational invariance in the asymptotic state. Finally, notice that off-diagonal correlations might influence the asymptotics of physically relevant observables: a simple example is  the asymptotic value of $\langle (M^{x}(t))^2 \rangle$, which for a generic non-translationally invariant condition $\mid \Psi_0 \rangle$ cannot be predicted using the GGE.

A very important open question is to understand under which general circumstances the GGE can be applied. For free fermionic and bosonic systems the GGE was argued to hold for local observables~\cite{cramer_08, barthel_08}. For more
general integrable systems this is not evident at all. For example in the case of the Quantum Ising chain the two significant observables, the transverse magnetization $\sigma^x_i$ and the order
parameter $\sigma^z_j$, are local in the spin representation. However, this locality does not translate directly to their representation in terms of the quasiparticles of the model: while
$\sigma^x_i$ retains a local character in terms of $\gamma_i$, $\sigma^z_i$ does not. Will  the asymptotic dynamics of any local operator be represented by the GGE, or just that of local operators in the quasi-particle fields ? Do the  symmetries of the initial state play any role in this ? Answering these questions appears to be crucial to understand the role of integrability in the dynamics of many-body systems.

Another very important question is whether all natural observables of an integrable system
behave necessarily nonthermally. The answer to it appears to be no, as  pointed out recently~\cite{rossini_09}. The key to understand this issue seems to be again \it locality \rm with respect to the quasi-particles diagonalizing the model. Thus in the Quantum Ising Model it was shown that while the transverse magentization is non-ergodic, the correlators of the order parameter $\sigma_z$ following a quench of the transverse field relax
as in a thermal state with an effective temperature $T_{\rm eff}$ set by the initial energy of the system $E=\langle \Psi_0(g_i) \mid H(g_f) \mid \Psi_0(g_i) \rangle$. At low $T_{\rm eff}$ this relaxation appears to be universal, i.e.
determined only by the low energy scattering properties of quasi-particles~\cite{rossini_09}. Analogous studies for an XXZ chain hint towards a different behavior of local and non-local operators with respect to  quasi-particles~\cite{canovi_10}. The situation is much less clear for quenches with high effective temperatures, where the universal character of the low energy theory is lost~\cite{barmettler}.

\subsection{Breaking integrability: eigenstate thermalization.}

When integrability is explicitly broken with a strong enough
perturbation one naturally expects ergodic behavior to emerge for
all observables~\cite{kollath_07, manmana, rigol_08, rigol_09,
roux_09, roux_10}. The quest for the necessary conditions for
thermalization to occur (i.e. how strongly should integrability be
broken, which spectral properties should the system display) is an
important problem in many different fields, from mathematical and
statistical physics to quantum chaos \cite{peres_84, deutsch_91,
srednicki_94, Srednicki1999, rigol_08}. In classical systems the
intense research on this subject was stimulated by the study of
dynamics of a nonlinear chain of coupled oscillators by Fermi Pasta
and Ulam (FPU)~\cite{fpu}, where instead of thermalization regular
quasi-periodic oscillations were observed. Later it was realized
that the FPU problem is nearly integrable and that there is a finite threshold for the chaotic behavior~\cite{campbell_05} . In quantum systems the situation was far less clear: while different views on this issue emerged from time to time, the key towards a
clear understanding of quantum thermalization appears to be linked
to the emergence of quantum chaotic behavior~\cite{peres_84}. In
particular, it has been proposed that the emergence of thermal
behavior is linked to the pseudo-random form of natural observables
once represented in the eigenbasis of the Hamiltonian~\cite{peres_84}. This observation has been made more
precise  by conjecturing that thermalization in quantum chaotic systems occurs eigenstate-by-eigenstate, i.e. the
expectation value of a natural observable $\langle \Psi_{\alpha} \mid A \mid \Psi_{\alpha} \rangle$ on an eigenstate $\mid \Psi_{\alpha} \rangle$ is a smooth function of its energy $E_{\alpha}$ being essentially constant on each microcanonical energy shell~\cite{deutsch_91, srednicki_94, Srednicki1999}. If this happens, then ergodicity and thermalization in the asymptotic state follow for every initial condition sufficiently narrow in energy
(e.g. localized in a microcanonical shell), as one can easily understand using the diagonal ensemble. This hypothesis is known as \it eigenstate thermalization \rm (ETH).

In order to understand how \it eigenstate thermalization \rm can emerge, let us consider a quantum gas of $N$ particles of mass $m$ with hard-core interactions~\cite{srednicki_94}. Srednicki pointed out that in the time evolution of this system starting with an initial condition $\mid \Psi_0 \rangle$ sufficiently narrow in energy, the momentum distribution will always relax to the Maxwell-Boltzmann distribution $f_{MB}(p)$ as long as the eigenstates of the system $\mid \Psi_{\alpha} \rangle$ can be considered as pseudo-random superpositions of plane waves, i.e. have
a diffusive nature in phase space. This requirement should be satisfied as a result of the chaoticity of the system, the so-called Berry's conjecture.  Calling ${\bf X}=({\bf x}_1,{\bf x}_2,,\dots, {\bf x}_N)$ the coordinates of the particles and ${\bf P}=({\bf p}_1,{\bf p}_2,\dots{\bf p}_N)$ their momenta,  Berry's conjecture
states that the eigenstates  have the form
\begin{eqnarray}
\Psi_{\alpha}({\bf X})={\cal N} \int d{\bf P} A_{\alpha}({\bf P})\delta({\bf P}^2-2m E_{\alpha}) e^{i{\bf P}\cdot {\bf X}},
\end{eqnarray}
with $A_{\alpha}({\bf P})$ being pseudo-random variables with gaussian statistics, $\langle A_{\alpha}({\bf P}) A_{\beta}({\bf P'}) \rangle=\delta_{\alpha}^{\beta}\delta^{(3N)}({\bf P}+{\bf P'})/\delta({\bf P}^2-{\bf P'}^2)$. Notice that we are disregarding the symmetrization of the wave function, see ~\cite{srednicki_94} for a discussion of this aspect. If the properties above are assumed it is easy to prove that on average in the thermodynamic limit  the momentum distribution function is
\begin{eqnarray}
\langle f ({\bf p})\rangle&=&\int d{\bf p}_2 d{\bf p}_3\;\dots \langle \mid \Psi_{\alpha}({\bf p},{\bf p_2},\dots) \mid^2 \rangle\nonumber \\
&=&\frac{e^{-\frac{{\bf p}^2}{2mkT}}}{(2\pi m k T)^{3/2}}=f_{MB}(p),
\end{eqnarray}
where the temperature is set by the equipartition law as $E_{\alpha}=3/2 N kT$. Notice that this is expected to happen for every eigenstate of energy close to $E_{\alpha}$, as required by the ETH. Hence thermal behavior will follow for every initial condition sufficiently narrow in energy.

For generic many-body systems, such as Hubbard-like models and spin chains, the close relation between breaking of integrability and quantum chaotic behavior is a known fact~\cite{poilblanc_93}. In particular, finite size many-body integrable systems are characterized by the Poisson spectral statistics while the gradual breaking of integrability by a perturbation leads to a crossover to the Wigner-Dyson statistics. The latter is typically
associated, in mesoscopic systems or billiards, with diffusive behavior and can be taken as a signature of quantum
chaos~\cite{imry_97}. In many-body disordered systems the emergence of the Wigner-Dyson statistics was argued to be an indicator of the transition between metallic (ergodic) and insulating (non-ergodic) phases~\cite{mukerjee_06,oganesyan_07}. Inspired by these close analogies, recent studies gave a boost to our understanding of the crossover from non-ergodic to thermal behavior as integrability is gradually broken and of the origin of ergodicity/thermalization in systems sufficiently far from integrability~\cite{kollath_07, manmana, rigol_08, biroli_10, rigol_09}. In particular, a careful study of the asymptotics of density-density correlators and momentum distribution function for hard-core bosons in $1d$ showed that the transition from non-thermal to thermal behavior \it  in finite size \rm systems takes the form of a crossover controlled by the strength of the integrability breaking perturbation and the system size~\cite{rigol_09}. Moreover there is a universality  in state to state fluctuations of simple observables in this crossover regime~\cite{neuenhahn_10}, which goes hand-by-hand with an analogous transition from   Poisson to  Wigner-Dyson level statistics~\cite{santos_10, rigol_10b}.  When integrability is broken by sufficiently strong perturbation ergodic behavior emerges~\cite{rigol_09, rigol_10b, neuenhahn_10}, which in turn appears to be related to the validity of the ETH~\cite{rigol_08}. In this context, the anomalous, non-ergodic behavior of integrable models has been reinterpreted as originating from wide fluctuations of the expectation value of natural observables around the microcanonical average~\cite{biroli_10}.

All these statements apply to the asymptotic (or time averaged)
state. So far the relaxation in time, in particular in the
thermodynamic limit, has received much less attention.  In a series
of studies of relaxation in fermionic Hubbard models subject to
quenches in the interactions strength it has been argued that for
sufficiently rapid quenches relaxation towards thermal equilibrium
occurs through a pre-thermalized phase~\cite{moeckel_08,
moeckel_10}. Similar two step dynamics occurs in quenches of coupled
superfluids where initial fast ``light cone'' dynamics leads to a
pre-thermalized steady state, which then slowly decays to the
thermal equilibrium through the vortex-antivortex
unbinding~\cite{mathey_10}. In Ref.~\cite{burkov_07} a very unusual
sub-exponential in time decay of correlation functions was predicted
and later observed experimentally~\cite{hofferberth_07} for
relaxational dynamics of decoupled 1d bosonic systems.

\subsection{Outlook and open problems:  quantum KAM threshold as a many-body delocalization transition ?}

The arguments above clearly pointed to the connection between
thermalization in strongly correlated systems and in chaotic
billiards. This analogy however , rather than being the end of a quest, opens an entire new kind of questions, which are a current focus of both theoretical and experimental research. In particular, we do know that in a number of models of strongly correlated particles eigenstate thermalization is at the
root of thermal behavior~\cite{rigol_08}. What is the cause of
eigenstate thermalization in a generic many-body system, i.e. the
analogue of the diffusive eigenstates in phase space of Berry's
conjecture ? And most importantly, while in a finite size system the
transition from non-ergodic to ergodic behavior takes the form of a
crossover, what happens in the thermodynamic limit ? Is the
transition from ergodic to non-ergodic behavior still a crossover or
it is sharp (a \it quantum KAM threshold \rm) ?

At present, research on these questions has just started. An interesting idea that has recently emerged is that  the
study of the transition from integrability to non-integrability in quantum many-body systems is deeply connected to  another important problem at the frontier of condensed matter physics:  the concept of many-body localization~\cite{altshuler_97,basko_06}, which extends the original work of Anderson on single-particle localization~\cite{anderson_58}. We note that related ideas were put forward in studying energy transfer in interacting harmonic systems in the context of large organic molecules~\cite{logan_90, leitner_96}. More specifically, it has been noticed that a transition from localized
to delocalized states either in real space~\cite{pal_10} or more
generally in quasi-particle space~\cite{canovi_10} is closely
connected to a corresponding transition from thermal to non-thermal
behavior in the asymptotics of significant observables. For weakly
perturbed integrable models, the main characteristic of the
observables to display such transition appears to be again their
locality with respect to the quasiparticles~\cite{canovi_10}.
This connection with many-body localization becomes
more clear on the basis of a recently proposed way to quantify the
transition from non-ergodic to ergodic behavior in many body
systems~\cite{olshanii_09}. The authors consider an integrable
Hamiltonian $H_0$ with a weak non-integrable perturbation $\lambda
V$. Formulating essentially a generalization of the Berry's
conjecture and making some additional assumptions they showed that
the deviations from thermal behavior in the expectation value of
observables can be quantified according to the formula
\begin{eqnarray}
\overline{\langle \Psi(t)\!\mid\!A\! \mid\! \Psi(t) \rangle}-\langle A \rangle_{\rm mc}
\approx \eta (\overline{\langle \Psi_0(t)\! \mid \!A\! \mid\! \Psi_0(t)\rangle_{0}}-\langle A \rangle_{\rm mc}), \nonumber
\end{eqnarray}
where $\mid \Psi_0(t) \rangle=\exp[-i{\cal H}_0 t] \mid \Psi_0 \rangle$ is
the time evolved state with respect to the integrable Hamiltonian,
while  $\mid \Psi(t) \rangle=\exp[-i({\cal H}_0+\lambda {\cal V}) t]
\mid \Psi_0 \rangle$ evolves with the non-integrable one.  The key ingredient in this formula is the parameter $\eta$, defined as the average over the microcanonical shell of the inverse participation ratio $\eta_{\alpha}=\sum_{n} \mid \langle \varphi_n \mid \Psi_{\alpha} \rangle \mid^4$, where $\mid \varphi_{n} \rangle$ are the eigenstates of the integrable Hamiltonian ${\cal H}_0$ and $\mid \Psi_{\alpha} \rangle $ are the eigenstates of ${\cal H}_0+\lambda {\cal V}$. Notice that when the system is close to integrability $\eta \simeq 1$ but as the strength of ${\cal V}$ increases, $\eta$  is roughly proportional to the inverse of the number of states ${\cal N}$ hybridized by the perturbation.

Using this formula  it is now possible to understand how many-body localization enters the scenario~\cite{pal_10,canovi_10}: an abrupt transition at a certain $\lambda_c$ from localized to delocalized
states in quasi-particle space corresponds to a sharp decrease of the inverse participation ratio $\eta$ from a value $O(1)$ to a value negligibly small and tending to zero in the thermodynamic limit, essentially $O(1/{\cal N}(\lambda))$, where ${\cal N}(\lambda)$ is the total number of states in an energy window of width of the order of the matrix elements of the perturbation. This would lead to an abrupt transition from non-thermal to thermal
behavior at $\lambda_c$, a transition qualitatively corresponding to the physics of a quantum KAM threshold. Notice that on the delocalized side of this transition the eigenstates are expected to be of diffusive nature (in quasi-particle space), i.e. a natural generalization of the form postulated by Berry's conjecture.
Such transition have been  studied extensively in confined electronic system, following a seminal paper by Althsuler, Kamenev, Levitov and Gefen~\cite{altshuler_97} and on interacting electron systems with localized single particle states~\cite{basko_06}. While the dependence $\eta(\lambda)$ was analyzed numerically in certain small systems~\cite{canovi_10, santos_10, santos_10a, neuenhahn_10}, eventual emergence of a sharp KAM-like threshold in the thermodynamic limit remains an open question.

The ETH also suggested a new way on looking at quantum relaxational
dynamics as dephasing in the many-body basis. In particular, the
information about the asymptotic state is fully contained in the
diagonal elements of the density matrix, which do not change in time
if the Hamiltonian is constant. So the process of time evolution in
this picture is equivalent to averaging of oscillating off-diagonal
elements of the density matrix to zero. In a way this picture is
different from conventional thinking based on kinetic theory of
thermalization through collisions between quasi-particles and the
time evolution of their distribution function. This apparent
difference is hidden in the complicated structure of the many-body
eigenstates. Our intuition is based on thinking about dynamics in
the approximate basis, e.g. a basis of independent quasi-particles.
The precise relation between the many-body and kinetic approaches to
thermalization is still an open question. Another potentially
intriguing possibility is understanding thermalization as a
renormalization group process, where time evolution results in
averaging over high-energy degrees of freedom. If one deals with
approximate noninteracting basis then because of interactions the
process of eliminating high energy states affects dynamics of low
energy modes and hence in renormalization of the low energy
dynamics. In Ref.~\cite{mathey_10} it was shown that such
renormalization process indeed can explain real time dynamics in a
two-dimensional sine Gordon model and the emerging nonequilibrium
Kosterlitz-Thouless transition. In Ref.~\cite{moeckel_09} similar
ideas were put forward to analyze dynamics of interacting fermions
using the flow equation method. At the moment it is unclear whether
using such real time renormalization group one can analyze
relaxational long time dynamics in generic interacting systems.

\section{Experimental progress in quantum dynamics in cold atoms and other systems}
\label{sec:exp}

As we mentioned before, the study of nonequilibrium
dynamics of quantum many-body systems has been increasingly
motivated by a series of advances in the field of ultracold atomic
and molecular gases. Due to the confluence of various features,
these mesoscopic quantum systems are in many ways near-ideal systems
for the study of nonequilibrium quantum phenomena.

Firstly, quantum gases can exhibit a remarkably high degree of
isolation from environmental sources of decoherence and dissipation.
Thus, to an excellent approximation, during duration of experiments they can be regarded as closed
quantum systems. Further, the dilute nature of these gases and
exceptionally low temperatures result in long timescales of
dynamical effects (typically on the order of milliseconds or longer)
allowing for time-resolved studies of nonequilibrium processes
resulting from phase-coherent many-body dynamics. Such studies are
hardly possible in conventional condensed matter systems.

Secondly, an array of techniques have been developed to dynamically tune various
parameters of the Hamiltonian governing these quantum gases. This has made it
possible to realize various prototypical nonequilibrium processes such as
quantum quenches discussed above. Quenches across phase transitions have been realized to
investigate the onset and formation of long range order and the mechanism
underlying the spontaneous formation of topological defects. The latter is
closely related to the KZ mechanism described earlier in the text. A quantitative experimental study of the
 defect, entropy and energy production resulting from such quantum quenches should allow for an accurate comparison with the theoretical predictions.

Lastly, the ability to engineer and experimentally realize model Hamiltonians
of archetypal correlated systems coupled with a detailed knowledge of the
microscopic interactions make ultracold atomic gases a tantalizing system
for applications ranging from the quantum simulation of strongly correlated
systems to the adiabatic quantum computation. In addition to the form of the
model Hamiltonian, experimental control can also be achieved over the
effective dimensionality of the ultracold gas making it possible to
investigate the nontrivial interplay between fluctuations, interactions
and dimensionality.

From a technological perspective, there is an increasing thrust towards engineering ultracold atomic many-body systems for applications in quantum metrology \cite{esteve, leroux, riedel, appel, meiser, veng1}. A deeper understanding of the dynamics of interacting many-body systems and the mechanisms of decoherence and
dissipation in these systems is of crucial importance in this context.

Motivated by these factors, a number of experiments have been
performed in recent years using ultracold quantum gases to
investigate topics including quantum coherent dynamics in optical
lattices, quenches across quantum phase transitions and
thermalization in low dimensional systems. For the purposes of this
colloquium, we distinguish between classes of nonequilibrium
experiments both in terms of the general protocol as well as the
questions being addressed by these experiments. (i) {\em
Nonequilibrium states} of many body atomic systems wherein the high
degree of isolation of the atomic system from the environment allows
for the creation of metastable or highly excited many-body states
with long lifetimes, (ii) {\em Quantum quench} experiments in which
one or more parameters of the Hamiltonian are changed rapidly to
create an out-of-equilibrium state of the many-body system and (iii)
{\em Dynamical tuning} of the Hamiltonian in order to study quantum
coherent dynamics of an interacting many-body system.

\begin{figure} \includegraphics[width=2in]{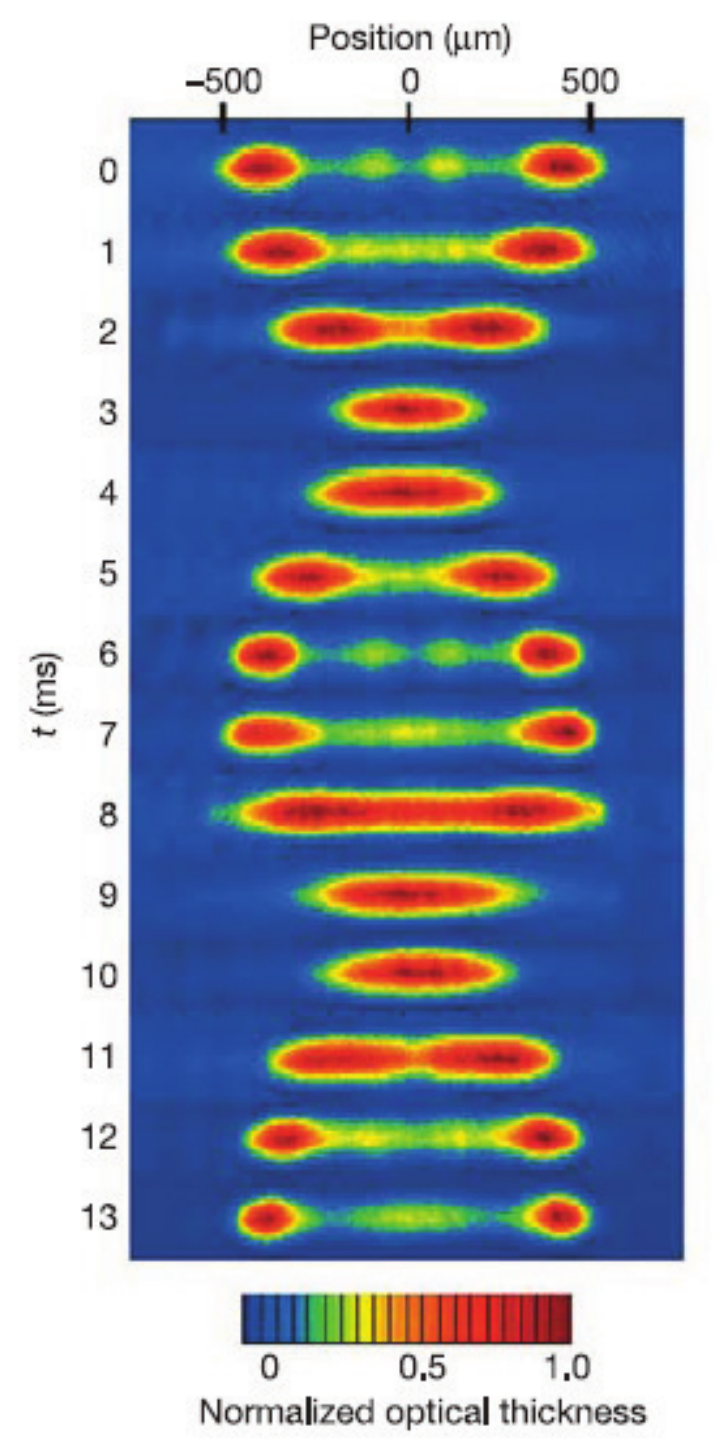}
\caption{{\bf Time-of-flight absorption images of an ensemble of
1D Bose gases} - Ultracold atoms are confined in arrays of 1D optical
traps. Optical pulses are used to place the atoms in a
superposition of $\pm 2 \hbar k$ momentum states. The gas is then
allowed to evolve for variable durations before being released from
the trap and photographed to reveal the momentum distribution.
The false color in each image is rescaled to show detail. The non-gaussian
nature of the momentum distribution clearly indicates an absence of
thermalization.({\em Adapted from \cite{kinoshita}})}
\label{fig:NC}
\end{figure}

These experimental advances have stimulated a very active
theoretical research in the area of nonequilibrium quantum dynamics
in interacting many-body systems. Among the issues most debated in
recent literature is the relation between thermalization in isolated
quantum systems and quantum integrability. In this regard, a recent
pioneering study on thermalization in 1D Bose gases was performed in
Ref.~\cite{kinoshita}. In this experiment, a blue detuned 2D optical
lattice was used to create arrays of tightly confined tubes of
ultracold $^{87}$Rb atoms. The depth of the lattice potential far
exceeded the energy of the ultracold gas ensuring negligible
tunneling among the tubes. The array of tubes was then placed in a
superposition of states of momentum $\pm 2 p_0$ by the application
of a transient optical phase grating. The imparted kinetic energy
was small compared to the energy required to excite the atoms to the
higher transverse states and the gases remained one-dimensional.
This out-of-equilibrium system was then allowed to evolve for
variable durations before the momentum distribution was probed by
absorption imaging of the gas (see Fig.~\ref{fig:NC}).

It was found that, while the initial momentum distributions exhibit
some dephasing on account of trap anharmonicities, the dephased
distribution remains non-gaussian even after thousands of
collisions. This is in distinct contrast to the gaussian
distributions observed when the 2D optical lattice is adiabatically
imposed on an equilibrium 3D Bose gas. This remarkable observation
that the nonequilibrium Bose gases do not equilibrate on the
timescales of the experiment appears consistent with the fact that
this system is a very close experimental realization of a
Lieb-Liniger gas with point-like collisional interactions - an
integrable quantum system in which only elastic pairwise
interactions can occur. Apparently the experimental technicalities
such as anharmonicities or the axial potential are insufficient to
sufficiently lift integrability in this system.

In addition to unambiguously showing the absence of thermalization within experimental
timescales in this model realization of the Lieb-Liniger gas, this study also
points the way towards addressing more general questions on integrability and
ergodicity. Starting from an integrable system, modifications such as the addition of finite range interactions, tunneling between the 1D tubes or the imposition of axial potentials one can tunably lift integrability and analyze emergence of irreversability and thermalization. This experiment largely motivated much of the theoretical work discussed in the previous section.

Another issue that has attracted a lot of attention is the search
for universal effects either in the nonequilibrium dynamics
following a quantum quench or in the adiabatic dynamics near a
quantum critical point which we described earlier in the text.
In particular, the issue of non-adiabatic dynamics near quantum phase transitions has been the focus of recent experimental studies on condensate formation in a dilute, weakly interacting Bose gas that is rapidly cooled past the BEC phase transition \cite{weiler}. This process was found to be accompanied by the spontaneous formation of topological defects, i.e. vortices, in the nascent superfluid. This can be
phenomenologically understood as being due to the formation of isolated
superfluid regions of a characteristic size $\xi$, each with a random relative
phase. These isolated regions then gradually merge to give rise to
global phase coherence. In this process, regions which enclose phase loops
of $2 \pi$ are constrained by the nature of the superfluid, i.e. the
continuity of the wavefunction, to have a vanishing superfluid density
at the core. Thus, the KZ mechanism predicts a
density of vortices that scales as $1/\xi^2$.

In this experiment, a magnetically trapped thermal gas of $^{87}$Rb atoms
was cooled by radiofrequency (rf) evaporation to temperatures below the
BEC transition temperature. The quench rate, i.e. the rate of cooling,
was controlled by varying the rate at which the rf frequency was ramped down.
Following a brief duration of equilibration, vortices are detected by
absorption imaging of the gas after ballistic expansion. Allowing for
some uncertainty in the ability to discern a vortex due to line-of-sight
integration in these images, it was found that about a quarter of the images
showed at least one vortex core.

The rate of cooling during the quench was limited
by the collision rate between atoms in the trapped gas during evaporative cooling.
This resulted in a limited dynamic range for the quench rate. Also,
the rapid decrease of the thermal fraction following the formation of the
condensate led to a low damping rate for the vortices.
A faster quench rate, realized through a trap with tighter confinement or
increased density or via sympathetic cooling with another species, could
result in the observation of an increased number of vortices during the quench.
In turn, this would potentially allow for quantitative tests of the predicted
scaling of vortex number with the quench rate and the extraction of dynamic
critical exponents.

While the formation of a superfluid by quenching the temperature is seeded by
thermal fluctuations, ultracold atomic gases also potentially allow for the
realization of phase transitions initiated purely by quantum fluctuations
 \cite{greiner2002a, sadler_06}.

 A particularly intriguing study of a quench past such quantum phase transitions was
 carried out in a degenerate $F=1$ spinor Bose gas of $^{87}$Rb \cite{sadler_06}.
 These gases, with a spin degree of freedom arising from a non-zero
 hyperfine spin $F$, are quantum fluids that may simultaneously exhibit
 the phenomena of magnetism and superfluidity, both of which result from
 symmetry breaking and long range order. Owing to rotational symmetry,
 the contact interactions between two atoms can be characterized by the total
 spin of the colliding pair. In the case of a $F=1$ spinor gas, these
 interactions give rise to a mean field energy given by $n (c_0 + c_2 \langle {\bf F} \rangle^2)$
 where the coupling strengths $c_{0,2}$ are related to the $s$-wave
 scattering lengths in the total spin $f=0,2$ channels~\cite{Ho1, Ohmi}.
 In addition to the mean field interactions, a finite external magnetic field $B$
 imposes a quadratic Zeeman energy (QZE) that scales as $q \langle F_z^2 \rangle$ with
 $q = (\mu_B B)^2/4 \Delta_{hf}$ where $\mu_B$ is the Bohr magneton and $\Delta_{hf}$ is
 the energy splitting between the ground state hyperfine manifolds.

 For a $F=1$ condensate of $^{87}$Rb,
 the competing influences of the spin-dependent interaction and the QZE give rise to
 a continuous quantum phase transition between a `polar' and a ferromagnetic phase.
 Rapidly tuning the external magnetic field from large values ($q \gg |c_2 n|$) to
 small values ($q \ll |c_2 n|$) quenches the spinor gas from the polar phase
 to the ferromagnetic phase. The ensuing growth of ferromagnetic domains was
 directly detected by {\em in situ} imaging (see Fig.~\ref{fig:KZ}). It was found that the resulting texture
 of ferromagnetic domains was spatially inhomogeneous can characterized by
 a typical length scale that was related to the spin healing length $\xi = \hbar/\sqrt{2 m |c_2 n|}$.
 Concurrent with the appearance of these domains, the spin textures revealed
 the spontaneous formation of polar-core spin vortices. These topological defects
 are characterized by a non-zero spin current but no mass current.
 The origin of these spin vortices is also rooted in the KZ mechanism. It was shown that,
 for slow quenches, the number of such vortices is expected to scale as $\tau^{-1/6}$
 where $\tau$ is the time over which the spinor gas is swept into the ferromagnetic
 state~\cite{saito}.

 The weak spin-dependent interactions inherent to this spinor gas also allow
 for nondestructive detection of the vortices and studies of their dynamics. In
 addition, the weak coupling between the spin and mass degrees of freedom make it straightforward to realize extremely low spin temperatures  to examine the role of quantum fluctuations in seeding this phase transition~\cite{klempt}.
 These features make spinor quantum fluids a rich system to investigate
 the quench dynamics and KZ mechanism past quantum phase transitions between
 different magnetically ordered phases.  In addition, corrections to the KZ scaling
 imposed by long range interactions~\cite{veng2}, conservation laws and finite
 temperature effects can also be studied.

Yet another range of experimental studies is made possible by the tunability of
atomic interactions using a Feshbach resonance. This technique allows the rapid
dynamic control of the $s$-wave scattering length by means of a time-varying
external magnetic field. This ability was utilized in a recent study of a strongly interacting two-component
Fermi mixture~\cite{jo}. Starting from an initially weak, repulsive interaction between the two Fermionic species, the interactions were rapidly increased by tuning the magnetic field to the vicinity of the Feshbach resonance. The
subsequent decrease in the atomic loss rate, the increase in the size of the trapped gas and the increase in kinetic energy as measured in time-of-flight images were interpreted as an indication of the Stoner transition
to a ferromagnetic state. However, in a later theoretical work this interpretation was questioned and an alternative explanation based on rapid molecule formation was suggested  (Ref.~\cite{babadi}). Thus, a direct in-situ measurement of local magnetization is necessary to understand whether ferromagnetism plays a role in this experiment.

In addition to the thermalization dynamics across phase transitions, the
long coherence times inherent to ultracold gases also makes it possible to
study the quantum coherent dynamics of many-particle systems. A particularly
dramatic instance of such coherent many-body dynamics was illustrated in the
collapse and revival of the matter wave field of a Bose condensate \cite{greiner2002b}.
Here, the interaction-induced dynamical evolution of a matter wave field
was clearly revealed in the multiple matter wave interference patterns
obtained after releasing the gas from the lattice. This work has also been
extended to the time-resolved observation of superexchange processes in
optical `superlattice' potentials \cite{trotzky_08}. Similar demonstrations of
collisional coherence have also been shown in spinor Bose gases \cite{chang,
kronjager}. Due to the internal degrees of freedom in a spinor gas, the
dynamics in this fluid is due to coherent spin-mixing collisions. In a
trapped gas that is well described by the single mode approximation (SMA),
these coherent collisions can lead to the periodic and reversible formation
of condensates in initially unpopulated spin states.

Further, in certain situations, coherent interactions can also lead to
quantum correlations~\cite{sorensen}. Schemes that might realize such entangled
many-particle states have received attention due to potential applications
in quantum information processing and metrology. The dynamical evolution
of such entangled states in the presence of quantum or thermal fluctuations
is obviously of great interest. A recent experiment investigated this
evolution in low dimensional two-component Bose gases with
adjustable interactions~\cite{widera}, finding that quantum fluctuations
play a crucial role in the phase diffusion dynamics of low dimensional
systems. More recently, the dynamical control of a Bose-Einstein condensate
confined in a strongly driven optical lattice was demonstrated~\cite{lignier}.
By periodically modulating the lattice potential, the tunneling parameter $J$ was shown to be suppressed in a phase coherent manner opening the possibility of driving quantum phase transitions using
this technique.

The isolation of ultracold atomic gases from external sources of dissipation also makes it possible to study
relaxation dynamics driven purely by intrinsic mechanisms. Such mechanisms should set the timescales for adiabatic quantum computing or the simulation of strongly correlated lattice models. A recent experiment along these lines
investigated the evolution of excited states of the repulsive Fermi-Hubbard system~\cite{strohmaier}. Here, doubly occupied lattice sites (doublons) were created by modulating the lattice and the subsequent decay of the system to
thermal equilibrium was monitored over time. It was shown that the lifetime of
these doublons scales exponentially with the ratio of the interaction energy to
kinetic energy, in fair agreement with theoretical predictions. It was argued that
the dominant mechanism driving this relaxation was a high-order scattering process
involving several fermions~\cite{sensarma_10}.

While this colloquium places an emphasis on experiments involving
ultracold atomic gases, there is a range of other mesoscopic quantum
systems which also lend themselves to studies on quantum
nonequilibrium dynamics. For the sake of completeness, we briefly
review a few of these systems here. Defect formation following a
quench was first studied in the context of vortices in liquid
crystals~\cite{chuang}. This has since been followed by similar
studies in various mesoscopic systems including isolated
superconducting loops where the defects assume the form of
spontaneous fluxoids~\cite{monaco_2009}, superconducting thin
films~\cite{maniv_2003} and multi-Josephson junction
loops~\cite{monaco_2002, monaco_2006}. A cumulative view of these
studies would indicate that the influence of finite size effects,
thermal fluctuations and dimensionality on the production of
topological defects by the KZ mechanism is as yet unclear and a
topic that warrants further study.

Another potential system for the study of nonequilibrium dynamics of
many-particle states arises from rapid advances in the field of
photonics. There have been several proposals~\cite{chang_2008,
greentree} for the dynamical creation of strongly correlated
photonic states using photon-photon interactions mediated by a
nonlinear optical medium. The realization of states such as a Tonks
gas of photons have been proposed using hollow-core optical fibers,
tapered optical fibers, photons in coupled cavities and surface
plasmons on conducting nanowires. Such strongly correlated photon
states should have applications in metrology, sub-shot noise
interferometry and the quantum emulation of exotic spin models.

\section{Outlook}

 One of the ultimate goals of the new field of quantum dynamics is to develop a systematic understanding of nonequilibrium phenomena in strongly interacting quantum many-body systems. A few of the most significant open questions along this avenue are readily identified : How can we classify nonequilibrium behavior in closed many-body systems? What is the general relation between integrability and dynamics? What is the dynamical effect of a gradual breaking of integrability? What are the effects of dissipation on these nonequilibrium processes? Can we understand time evolution of interacting systems through the renormalization group? Answering these and other questions allied with systematic, quantitative studies of possible nonequilibrium quantum phase transitions and the extraction of dynamical critical exponents, are just a few of the many tantalizing programs to be pursued. The rapidly developing sophistication and precision of ultracold atomic experiments and other experimental systems should allow for close and direct comparison between theoretical predictions and {\em ad hoc} experiments.

The realization of robust techniques for the experimental study of such systems and the development of theoretical tools to describe nonequilibrium many-body processes should bode for tantalizing opportunities in this nascent field, potentially leading to a deeper understanding of the principles governing nonequilibrium many-body phenomena and establishing robust connections between microscopic dynamics and statistical physics.

\section{acknowledgements}


A.P. was supported by NSF (DMR-0907039), AFOSR FA9550-10-1-0110 and the Sloan Foundation. K.S. thanks DST, India for financial support under Project No. SR/S2/CMP-001/2009. M. V. was supported by the Sloan Foundation.

\bibliography{NonequilRMP}
\end{document}